\documentclass[twocolumn,showpacs,amsmath,amssymb,superscriptaddress,floatfix]{revtex4-2}
\usepackage[breaklinks=true,colorlinks,citecolor=blue,linkcolor=blue,urlcolor=blue]{hyperref}
\usepackage[T1]{fontenc}
\usepackage{lineno}
\usepackage{textcomp}
\usepackage{amsmath}
\usepackage{graphicx}
\usepackage{multirow}
\usepackage{makecell}
\usepackage{booktabs}
\usepackage{derivative}
\usepackage{bbold}
\usepackage{lipsum}
\usepackage{array}
\usepackage{graphicx}
\usepackage{xcolor}
\usepackage{subfigure}
\usepackage{hyperref,hypcap}
\usepackage{braket}
\usepackage{amsmath}
\usepackage{commath}
\usepackage{mathtools}
\usepackage{chemformula}
\usepackage{natbib}
\usepackage{comment}
\newcolumntype{P}[1]{>{\centering\arraybackslash}p{#1}}
\usepackage{orcidlink}

\def\be{\begin{equation}}
	\def\ee{\end{equation}}

\def\bea{\begin{eqnarray}}
	\def\eea{\end{eqnarray}}

\makeatletter
\usepackage{lipsum}
\usepackage{epsfig}

\usepackage[colorlinks,linkcolor=blue]{hyperref}
\usepackage[nameinlink,capitalise]{cleveref}

\usepackage{subfigure}
\usepackage{color}
\usepackage{physics}

\definecolor{Red}{rgb}{1,0,0}
\definecolor{Blu}{rgb}{0,0,1}
\definecolor{Green}{rgb}{0,1,0}

\makeatother
\usepackage{amssymb}
\usepackage[english]{babel}

\begin{document}
	
	\title{Spin-orbit torque switching of N\'eel order in band-inverted antiferromagnetic bilayer MnBi$_2$Te$_4$}
	\author{Rajibul Islam\orcidlink{0000-0001-9076-7843}}
	\affiliation{Department of Physics, University of Alabama at Birmingham, Birmingham 35294, AL, USA}
	
	\author{Shakeel Ahmad\orcidlink{0009-0000-4994-9717}}
	\affiliation{Department of Physics, University of Alabama at Birmingham, Birmingham 35294, AL, USA}
	
	\author{Fei Xue\,\orcidlink{0000-0002-1737-2332}} 
	\affiliation{Department of Physics, University of Alabama at Birmingham, Birmingham 35294, AL, USA}

	\begin{abstract}
		Magnetic topological insulators host exotic phenomena such as the quantum anomalous Hall effect and quantized magnetoelectric responses, but dynamic electrical control of their topological phases remains elusive. Here we demonstrate from first principles that spin–orbit torque enables direct electrical switching of the Néel configuration in intrinsic antiferromagnetic bilayer \ch{MnBi2Te4}, thereby reconfiguring its boundary spectrum. A symmetry-allowed interband (time-reversal even) torque persists inside the bulk gap, and deterministically reverses the Néel order and layer-resolved Chern marker without free carriers.  Upon doping, both interband and intraband torques are amplified, lowering the critical electric field for switching by two orders of magnitude. Together, these results establish two complementary regimes of control: dissipationless in-gap torques without Joule heating and enhanced current-induced torques, providing a robust route to manipulate layer-resolved Chern marker and helical-like gapped edge modes in antiferromagnetic \ch{MnBi2Te4}.
	\end{abstract}
	\date{\today}
	\maketitle
	
	\section{Introduction}
	Magnetic topological materials provide a fertile platform for realizing exotic quantum phenomena, including quantum anomalous Hall effect, axion electrodynamics, and quantized magnetoelectric responses~\cite{Hasan2010,Qi2011,Tokura2019,Bernevig2022}. A central challenge, however, is to achieve dynamical control of these topological states. To date, most demonstrations of topological phase transitions have relied on static tuning parameters, such as external magnetic fields, chemical substitution, or structural modifications including layer thickness and twist angle~\cite{Chang2013,Kou2015,Kawamura2018,Serlin2020,Chen2020,Wu2020,Zhao2020,Li2021,Cai2023}. By contrast, electrical control would enable fast, reversible, and scalable manipulation of topology, a prerequisite for device applications~\cite{Chang2024}.
	In this context, the recent discovery of the intrinsic antiferromagnetic topological insulator~\cite{Mong2010} \ch{MnBi2Te4} offers a particularly promising material platform, where efficient manipulation of the layer magnetization directly tunes the underlying topological phase~\cite{Otrokov2019,Li2019,Wang2019_Axion,Otrokov2019prl,Deng2020,Liu2020,Klimovskikh2020,He2020,Lei2020,Gao2021,Zhao2021,Xie2023}. Establishing practical routes for electric control is therefore a pressing goal.
	
	Spin–orbit torque (SOT) provides one such route. In general, SOT arises in systems lacking inversion symmetry and enables  current- or electric-field-driven reorientation of magnetic order mediated by spin–orbit coupling~\cite{Manchon2019review,Shao2021}. It has been extensively studied in ferromagnets~\cite{miron2011perpendicular,liu2012spin,liu2012current,Kurebayashi2014} and compensated antiferromagnets~\cite{Zelezny2014,wadley2016electrical,Jungwirth2016,Mn2Au_Experiment2018,Baltz2018}, as well as in magnetic topological heterostructures, where surface states provide efficient current-driven torques that switch adjacent ferromagnets~\cite{Mellnik2014,Fan2014,Yang2015,Fan2016,Han2017,Wu2019,Tai2024}. In these cases topology mainly enhances the torque efficiency, while the topological phase itself remains fixed. By contrast, in intrinsic antiferromagnetic topological insulators such as \ch{MnBi2Te4}, magnetic order and band topology are inseparably linked. Additionally, although the bulk \ch{MnBi2Te4} respects $\mathcal{PT}$ symmetry, \textit{local} inversion asymmetry at each magnetic sublattice allows both staggered and uniform SOT components under in-plane electric fields~\cite{Zelezny2014, Xue2021}, making SOT a natural knob to directly reconfigure the magnetic configuration and associated boundary properties.
	
	Recent experiments on magnetically doped Chern insulators demonstrated that SOT-driven topological switching, but only with the aid of an external in-plane magnetic field to break mirror symmetry and enable deterministic reversal~\cite{Chang2024}. By contrast, intrinsic antiferromagnetic topological insulators such as \ch{MnBi2Te4} naturally avoid this limitation, possessing only a single in-plane mirror symmetry, similar to the bilayer \ch{CrI3} case~\cite{Xue2021}. On the theory side, Tang and Cheng~\cite{Cheng2024} have predicted dissipationless in-gap torques~\cite{Hanke2017} and associated charge conversion in \ch{MnBi2Te4} using an effective model. While effective models have offered qualitative insights~\cite{Cheng2024}, they often neglect the complex multiband hybridization and orbital texture essential for accurate torque prediction in real materials. Here, we provide a quantitative first-principles evaluation of the torque, resolving its full angular dependence under magnetic symmetry constraints, and demonstrating deterministic sublattice-resolved Néel switching in bilayer \ch{MnBi2Te4}. Our results show that the in-gap interband (time-reversal–even) torque is symmetry-allowed, persists throughout the bulk gap, and can reversibly switch the Néel order and the associated layer-resolved Chern marker without auxiliary magnetic fields. Moreover, we demonstrate that doping amplifies both interband and intraband contributions, providing an additional regime of efficient electrical control.
	
	The remainder of this paper is organized as follows. In Sec.~\ref{sec:crystal} we summarize the crystal and magnetic symmetries of bilayer \ch{MnBi2Te4} and the resulting constraints on uniform and staggered torque components. In Sec.~\ref{sec:bands} we present first-principles results for the band structure and the chemical-potential dependence of the sublattice-resolved torkance, including the finite interband response in the gapped regime. In Sec.~\ref{sec:topology} we analyze the topological characteristics and present the helical-like edge spectrum. In Sec.~\ref{sec:dynamics} we parameterize the angular dependence using the VSH expansion and perform LLG simulations demonstrating deterministic N\'eel switching in both insulating and doped regimes. In Sec.~\ref{sec:conclusion}, we summarize our findings and discuss the experimental relevance. Additional computational details and supporting analyses are provided in the Appendices.
	
	\begin{figure}[htbp]
		\includegraphics[width=0.9\linewidth]{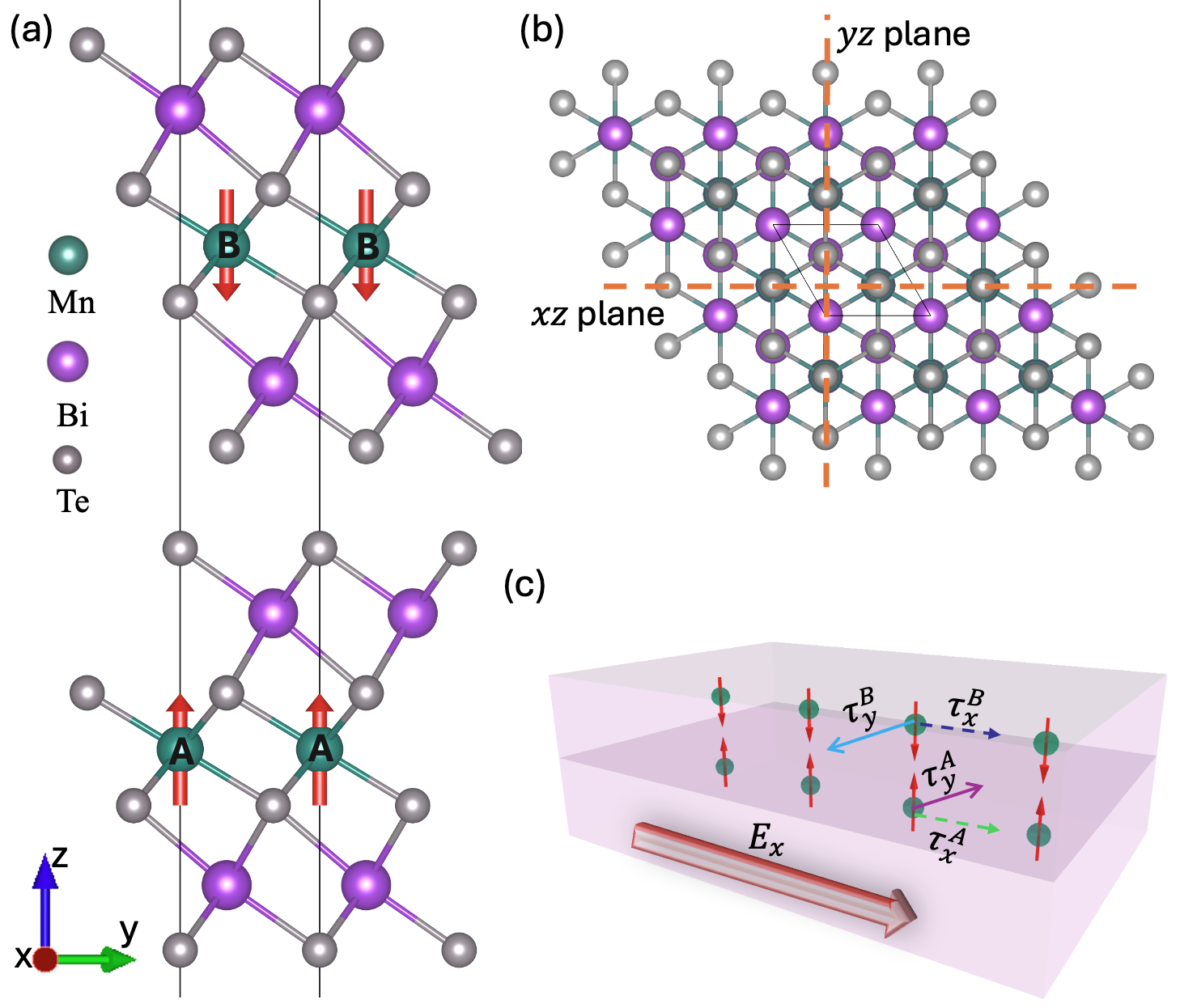}
		\caption{(a) Side and (b) top views of bilayer MnBi$_2$Te$_4$ with $\mathcal{PT}$-symmetric antiferromagnetic (AFM) order (red arrows indicate Mn moments). Each Mn site lacks local inversion symmetry, while the two Mn sublattices A and B are related by $C_{2x}$ and by $\mathcal{PT}$. Dashed lines in (b) mark the $yz$ and $xz$ planes; only the $yz$ plane is a mirror ($\mathcal{M}_{yz}$ present, $\mathcal{M}_{xz}$ broken). (c) Symmetry-allowed spin-orbit torques for an in-plane field $E_x$: transverse components ($\tau_y,\tau_z$) are staggered between layers, while the longitudinal component ($\tau_x$) is uniform. For brevity, we denote $\tau_i \equiv \tau_{ix}$.}
		\label{fig1: crystal}
	\end{figure}
	
	\section{Crystal symmetries and allowed spin-orbit torques}\label{sec:crystal}
	Bulk MnBi$_2$Te$_4$ crystallizes in a rhombohedral layered structure (space group $R\bar{3}m$) with out-of-plane easy axis and interlayer antiferromagnetic coupling~\cite{Li2019,Otrokov2019,Cao2020,Xie2023}. As shown in Fig.~\ref{fig1: crystal}(a), each septuple layer consists of Te–Bi–Te–Mn–Te–Bi–Te stacking. For the bilayer AFM ground state with $\hat{\mathbf{L}}=\hat{\mathbf{m}}^{\rm A}=-\hat{\mathbf{m}}^{\rm B}\parallel\hat{\mathbf{z}}$, the two septuple layers (Mn sublattices A/B) are related by $C_{2x}$ and by $\mathcal{PT}$ symmetries. Each layer has $C_{3z}$ and three symmetry-equivalent vertical mirror planes. In Fig.~\ref{fig1: crystal}(b) we choose coordinates such that one of them is $\mathcal{M}_{yz}$; the other two are obtained by $C_{3z}$ rotations. In this convention $\mathcal{M}_{xz}$ is
	not a mirror symmetry of the layer. For $\hat{\mathbf{L}}\parallel\hat{\mathbf{z}}$ this yields magnetic point group $\bar{3}m'$. The magnetic symmetry may be written using $\{C_{3z},\,\mathcal{PT},\,\mathcal{M}_{yz}\mathcal{T}\}$; since $(\mathcal{M}_{yz}\mathcal{T})(\mathcal{PT})=C_{2x}$, we equivalently use $\{C_{3z},\,\mathcal{PT},\,C_{2x}\}$ to make the A/B sublattice exchange explicit.

	These symmetries constrain the linear SOT on sublattice $\alpha\in\{\rm {A,B}\}, T_i^\alpha=\tau_{ij}^\alpha E_j$. Local inversion breaking at the Mn sites permits nonzero sublattice torques, while crystal symmetries, acting jointly on real space and spin in the presence of spin-orbit coupling (SOC), relate $\tau^{\rm A}$ and $\tau^{\rm B}$ [Fig.\ref{fig1: crystal}(c)]. For $\mathbf{E}\parallel\hat{x}$, the field is invariant under $C_{2x}:(x,y,z)\rightarrow(x,-y,-z)$, while $C_{2x}$ exchanges A/B; consequently, components that are even (odd) under this operation produce uniform (staggered) responses, giving uniform $\tau_{xx}$ but staggered $\tau_{yx}$ and $\tau_{zx}$. In addition, $\mathcal{PT}$ requires $\mathcal{PT}$-even (odd) torques to be staggered (uniform) across sublattices~\cite{Zelezny2014,Zelezny2017}. While both $C_{2x}$ and $\mathcal{PT}$ characterize the collinear AFM ground state, the subsequent dynamics generally develops a uniform canting component (from anisotropy and uniform torque channels) that breaks $\mathcal{PT}$, whereas the $C_{2x}$ sublattice-exchange relation remains the key constraint we use below to reduce the parameter space.

	\begin{figure}[htbp]
		\includegraphics[width=0.95\linewidth]{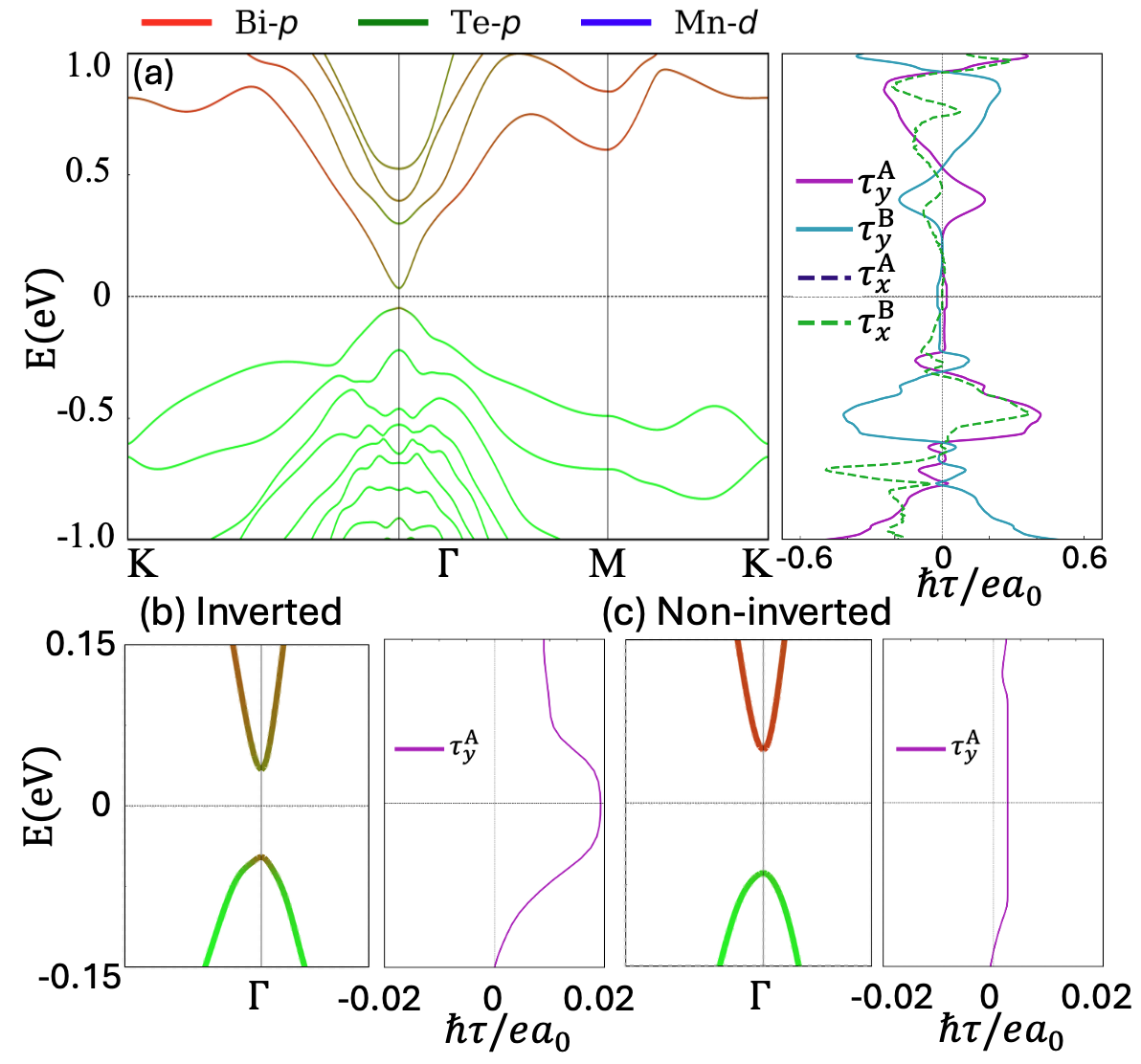}
		\caption{(a) Atom-resolved band structure of bilayer \ch{MnBi2Te4} (left) showing inverted Bi–$p$ and Te–$p$ character near the Fermi level, together with the sublattice-resolved spin–orbit torkance versus chemical potential (right).  The symmetry-enforced structure is evident: $\tau_{x}^{\rm A}=\tau_{x}^{\rm B}$ (uniform) and $\tau_{y}^{\rm A}=-\tau_{y}^{\rm B}$ (staggered). (b) Enlarged view of the inverted phase: the band structure (left) and the staggered torque $\tau_{y}^{\rm A}$ (right) highlight its finite in-gap response. (c) Same analysis with reduced spin–orbit coupling, showing the non-inverted bands where the in-gap torque is strongly suppressed.}
		\label{fig2: bands} 
	\end{figure}
	
	\section{Band Inversion Enhanced Spin-Orbit Torque in the Gap}\label{sec:bands}
	Having established the symmetry constraints governing spin–orbit torques in bilayer antiferromagnetic \ch{MnBi2Te4}, we now turn to microscopic calculations of the band structure and torque response. We perform first-principles calculations using the VASP package~\cite{VASP}, incorporating self-consistent SOC and van der Waals corrections, as detailed in the Appendix~\ref{Appendix:methods}. The SOTs are computed via linear response theory using a Wannier-interpolated tight-binding model~\cite{Wannier90}, as detailed in the Appendix~\ref{appendix:sot}. 
	
	Figure~\ref{fig2: bands}(a) shows the element-resolved band structure, revealing an $\sim$80~meV gap and a clear band inversion between Bi-$p$ and Te-$p$ orbitals at the $\Gamma$ point. Because the ground state preserves $\mathcal{PT}$ symmetry (with $\hat{\mathbf{L}}\parallel\hat{z}$), all bands remain doubly
	degenerate.
	The right panel of Fig.~\ref{fig2: bands}(a) plots the corresponding
	spin–orbit torkance under $E_x$. Consistent with the $C_{2x}$ symmetry analysis, $\tau_{yx}$ is staggered between sublattices, whereas $\tau_{xx}$ is uniform.  Furthermore, the two components arise from distinct symmetry channels: $\tau_{yx}$ originates purely from the time-reversal–even (interband) contribution, while $\tau_{xx}$ is purely time-reversal–odd (intraband). This distinction follows from the sublattice symmetry $\mathcal{M}_{yz}\mathcal{T}$: since both $T_y$ and $E_x$ are odd under $\mathcal{M}_{yz}\mathcal{T}$, $\tau_{yx}$ must be even under time reversal, whereas $T_x$ is even, so $\tau_{xx}$ originates solely from the odd channel. (Equivalently, $\mathcal{PT}$ symmetry enforces the same channel separation: $\mathcal{T}$-even responses are staggered between A/B while $\mathcal{T}$-odd responses are uniform~\cite{Zelezny2014,Xue2021}.)
	This symmetry-enforced distinction has a key implication: the odd (intraband) contribution vanishes in the gapped regime due to the absence of Fermi-surface states, whereas the even (interband) contribution can remain finite and thus dominates the torque response inside the gap. This is precisely what we obtain in our \textit{ab initio} torkance calculations:  the staggered even $\tau_{yx}$ (cyan and purple lines) survives inside the gap [see zoom-in view in Fig.~\ref{fig2: bands}(b)] whereas the uniform odd $\tau_{xx}$ (black and green dashed lines) vanishes. 
	
	Although smaller than the peak values outside the gap, the in-gap even torkance is strikingly large in absolute terms. Its magnitude reaches  $0.02\,ea_0/\hbar$, roughly $360$ times larger than the topological magnetoelectric quantum $\alpha_Q=\mu_0 e^2/(2h)=5.3\times 10^{-5}\,ea_0/(\hbar\gamma)$~\cite{TME2015}, where $\mu_0$ is the vacuum permeability constant, $\gamma$ is the gyromagnetic ratio, and $a_0$ is Bohr radius. In experimental units, this corresponds to $8.9\times10^{-5}$~Oe·m/V, in line with effective-model estimates for multilayer \ch{MnBi2Te4}~\cite{Cheng2024}. 
	
	To test the role of band inversion, we artificially reduced the spin–orbit coupling in VASP. As shown in Fig.\ref{fig2: bands}(c), the band inversion at $\Gamma$ disappears and the gap widens to $\sim$170 meV. Because interband contributions scale as $1/\Delta E^2$, we first rescaled the denominators to match the effective gap of the inverted case. Even then, the non-inverted system produces a torque nearly an order of magnitude smaller. Using the actual ($170$ meV) non-inverted gap without rescaling further suppresses the even torkance, by nearly a factor of $20$. This enhancement originates from the inverted dispersion, where conduction and valence bands remain nearly degenerate over extended $k$-space regions, creating interband “hot spots” that strongly amplify the torque. As shown in the Appendix~\ref{Appendix:model}, a similar band-inversion-enhanced response also arises in a minimal Chern insulator model~\cite{Qi2006}, indicating that this amplification mechanism is generic to systems with inverted bands.
	
	\begin{figure}[htbp]
		\includegraphics[width=0.9\linewidth]{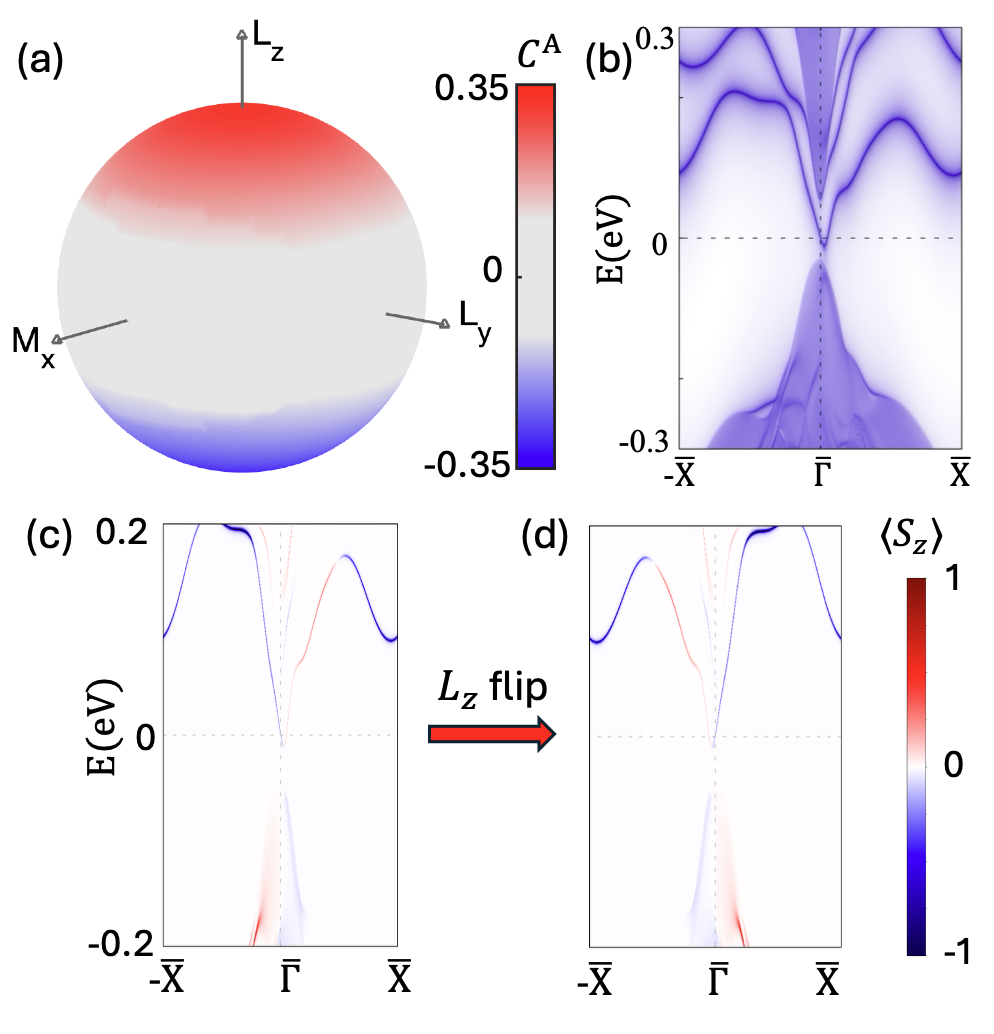}
		\caption{(a) Angular dependence of the sublattice-resolved local Chern marker $C^{\rm A}$; by $C_{2x}$ symmetry the opposite sublattice satisfies $C^{\rm B}=-C^{\rm A}$, so the total Chern number vanishes. 
			(b) Edge spectrum showing in-gap edge states. 
			(c,d) Zoomed-in edge dispersions, color-coded by spin expectation $\langle S_z\rangle$, 
			demonstrating quasi-helical spin textures that reverse their polarization upon 
			flipping the Néel vector $L_z$.}
		\label{fig3: topology}
	\end{figure}
	
	\section{Topological Characterization}\label{sec:topology}
	To further clarify the role of band inversion, we examine the topological characteristics of the bilayer system. Because of $\mathcal{PT}$ and $C_{2x}$ symmetries, the total Chern number vanishes, precluding a quantized anomalous Hall effect. Nevertheless, the inverted bilayer hosts nontrivial symmetry-based topology beyond the first Chern number. As shown in Appendix~\ref{Appendix:methods}, it realizes a $C_{3z}$-protected higher-order topological
	insulator characterized by rotation topological invariants (RTIs)
	extracted from the $C_{3z}$ eigenvalues of the occupied bands at the
	rotation-invariant momenta $\Gamma$, $K$, and $K'$~\cite{Wang2024,Hughes2019,Murakami2021}.
	For the inverted bilayer we obtain
	$\chi^{(3)}=\{[K_1^{(3)}],[K_1^{\prime(3)}],[K_2^{(3)}],[K_2^{\prime(3)}]\}
	= \{1,1,0,-1\}$, consistent with the band-inverted phase.
	
	Additionally, the layer-resolved Chern marker can be finite, even when the total Chern number vanishes. In the limit of weak interlayer
	hybridization, it approaches the half-quantized value expected for an
	axion-insulator thin film with gapped Dirac surface states~\cite{Vanderbilt2018}.
	As shown in Fig.~\ref{fig3: topology}(a), the inverted phase exhibits a sizable layer Chern marker. We display $C^{\rm A}$ for one sublattice, while the opposite $C^{\rm B}=-C^{\rm A}$ is explicitly obtained everywhere on the sphere, consistent with $C_{2x}$ symmetry. At $\hat{\mathbf{m}}^{\rm A}=-\hat{\mathbf{m}}^{\rm B}=\hat{\mathbf{z}}$, $C^{\rm A}$ is close to but not exactly $1/2$ due to hybridization effects. In contrast, the non-inverted phase obtained by reducing SOC yields a strongly suppressed layer Chern marker. Flipping the Néel order to $\hat{\mathbf{m}}^{\rm A}=-\hat{\mathbf{m}}^{\rm B}=-\hat{\mathbf{z}}$ reverses the sign of $C^{\rm A}$, demonstrating direct electrical control of the layer-resolved Chern marker via Néel switching
	
	The edge spectrum [Fig.\ref{fig3: topology}(b–d)] shows in-gap
	spin-textured edge modes whose spin polarization reverses upon
	flipping $L_z$. These states resemble the helical modes of a quantum spin Hall insulator. However, they acquire a small gap due to broken time-reversal symmetry, consistent with earlier studies of magnetically doped topological insulators~\cite{Zhang2013} and even-layer \ch{MnBi2Te4}~\cite{Lin2022}. By contrast, in the non-inverted regime the edge spectrum shows only weak boundary features without a comparable spin-textured structure (Fig.~\ref{Fig:edge} in Appendix~\ref{Appendix:methods}).
	
	Together, the Néel-controlled sign change of the layer Chern marker and
	the reversal of the edge-mode spin texture provide a direct manifestation of how magnetic switching reconfigures the
	boundary electronic structure in the band-inverted phase, an effect we
	exploit below in the SOT-driven spin dynamics.

	\begin{figure}[htbp]
		\includegraphics[width=0.95\linewidth]{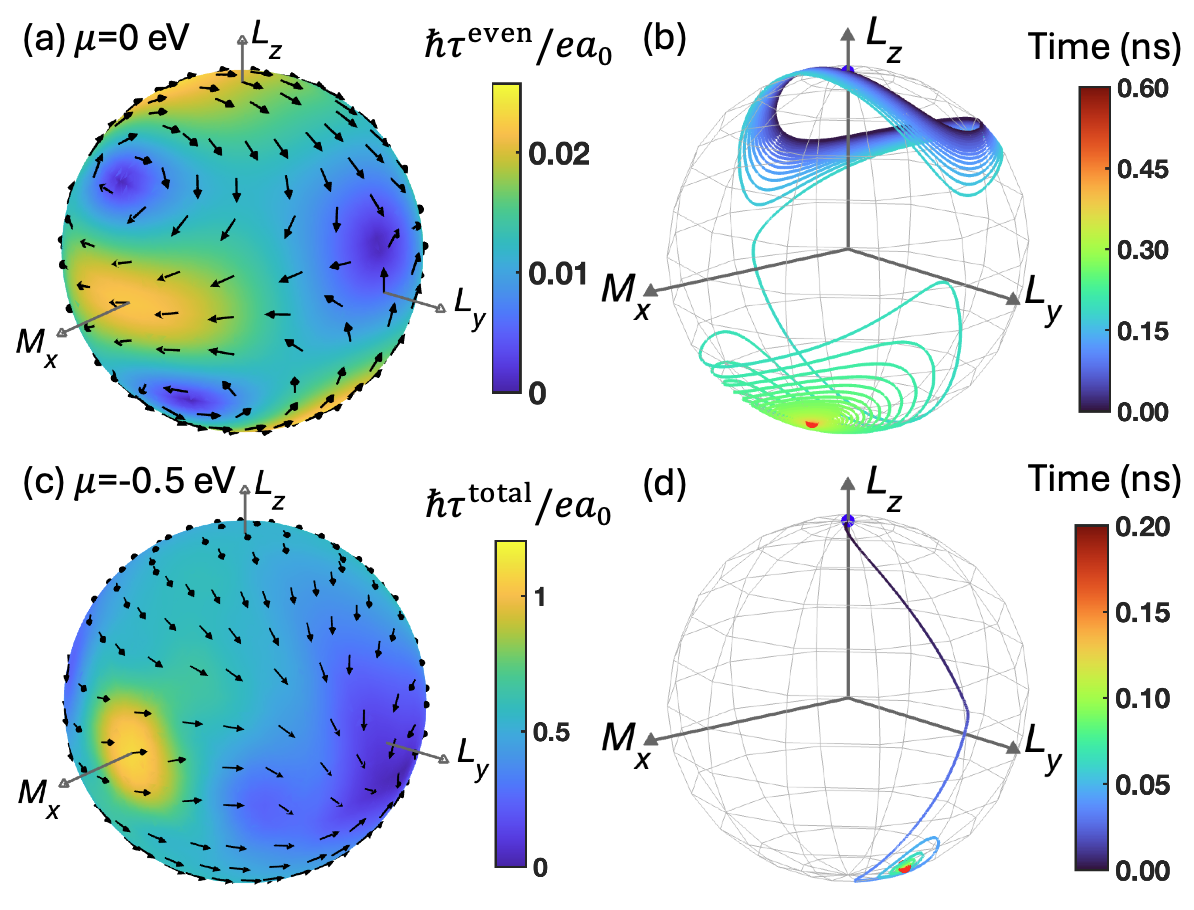}
		\caption{(a) Angular dependence of the even interband torkance $\hbar\tau^{\rm even}/ea_0$ at $\mu=0$ (in-gap), shown on the mixed vector sphere with arrows indicating the torque direction and color denoting its magnitude.
			(b) Simulated mixed vector dynamics under this torkance with critical electric field $0.32~\rm{V/nm}$, revealing deterministic switching between $\pm L_z$ on a sub-ns timescale.
			(c) Total torkance $\hbar\tau^{\rm total}/ea_0$ including both time-reversal even interband and odd intraband contributions at $\mu=-0.5$ eV.
			(d) Corresponding mixed vector dynamics at much smaller critical electric field $2.32$ V/$\mu$m (two orders of magnitude lower), showing slightly faster dynamics into the $-L_z$ state.}
		\label{fig4: spindynamics}
	\end{figure}

	\section{Spin Dynamics and N\'eel Order Switching} \label{sec:dynamics}
	Having established the enhanced in-gap time-reversal even staggered torque in the band-inverted bilayer \ch{MnBi2Te4}, we now analyze its angular dependence and impact for antiferromagnetic dynamics. The sublattice magnetizations evolve according to coupled Landau–Lifshitz–Gilbert equations with additional electric-field-induced spin–orbit torque terms~\cite{SLONCZEWSKI1996,Gomonay2010,Baltz2018,Xue2021} 
	\begin{equation}
		\label{eq:LLG}
		\frac{d{\hat{\mathbf{m}}}^{\rm A,B}}{dt}=\hat{\mathbf{m}}^{\rm A,B} \times \left(\frac{\gamma}{m}\frac{\delta E}{\delta \hat{\mathbf{m}}^{\rm A,B}}  +\alpha~\frac{d{\hat{\mathbf{m}}}^{\rm A,B}}{dt}\right)+\boldsymbol{\mathcal{T}}^{\rm A,B}\\
	\end{equation}
	where $m$ is the sublattice moment magnitude (assumed equal on both sublattices) and $\alpha$ is the Gilbert damping parameter. The energy functional $E$ includes easy-axis anisotropy (along $\hat{\bf z}$) and Heisenberg exchange coupling: 
	\begin{equation}
		\label{eq: energy}
		E=-\frac{1}{2}m K[\left(\hat{\mathbf{m}}^{\rm A}\cdot\hat{\bf z}\right)^2+\left(\hat{\mathbf{m}}^{\rm B}\cdot\hat{\bf z}\right)^2 ] + m H_{\rm E}\left(\hat{\mathbf{m}}^{\rm A}\cdot\hat{\mathbf{m}}^{\rm B}\right)
	\end{equation} 
	where $K$ and $H_{\rm E}$ are the effective magnetic fields from anisotropy and exchange, respectively.
	
	In principle, the torque must be resolved in the enlarged parameter space of four angular degrees of freedom $(\theta^{\rm A/B},\phi^{\rm A/B})$ due to the exchange term in Eq.~\ref{eq: energy}. Such an $N^4$ space makes first-principles sampling intractable. However, the system retains $C_{2x}$ symmetry when the initial state satisfies it. This symmetry enforces $T_x^{\rm A}=T_x^{\rm B}$ and $T_{y,z}^{\rm A}=-T_{y,z}^{\rm B}$ for all torque contributions: exchange, anisotropy, and spin–orbit. As a result, the dynamics preserve the uniform character of the $x$ component and the staggered character of the $y$ and $z$ components. We therefore can replace the full $(\mathbf{m}^{\rm A},\mathbf{m}^{\rm B})$ description by a reduced mixed vector representation,
	$\mathbf{N}\equiv(M_x,L_y,L_z)=\tfrac{1}{2}(\hat{m}_x^{\rm A}+\hat{m}_x^{\rm B},\, \hat{m}_y^{\rm A}-\hat{m}_y^{\rm B},\, \hat{m}_z^{\rm A}-\hat{m}_z^{\rm B}$),
	which captures the dynamics in a two-dimensional angular space, analogous to a ferromagnet. 
	Notably, while the ground state also preserves $\mathcal{PT}$ symmetry, the easy-axis anisotropy torque and uniform spin-orbit torque immediately introduce canting that breaks it~\cite{Zelezny2014, Xue2021}. By contrast,  $C_{2x}$ symmetry remains robust and ensures that the staggered/uniform decomposition of torques holds throughout the dynamics. This construction closely parallels our earlier treatment of \ch{CrI3} under $C_{2y}$ symmetry~\cite{Xue2021}.

	To probe the angular dependence explicitly, we rotate the magnetization on sublattice A by angles $(\theta^{\rm A},\phi^{\rm A})$, while enforcing $C_{2x}$ symmetry for sublattice B, {\it i.e.} $\theta^{\rm B}=\pi+\theta^{\rm A}$ and $\phi^{\rm B}=\pi-\phi^{\rm A}$. This procedure ensures that the torque components obey the correct symmetry relations ($T_x^{\rm A}=+T_x^{\rm B}$, $T_{y,z}^{\rm A}=-T_{y,z}^{\rm B}$), and allows the angular dependence to be efficiently mapped onto the reduced mixed vector $\mathbf{N}=(M_x,L_y,L_z)$ using only $(\theta^{\rm A},\phi^{\rm A})$.
	The resulting torque distribution [Fig.~\ref{fig4: spindynamics}(a)] exhibits pronounced higher-order angular dependence beyond the canonical lowest-order form, with both dampinglike and fieldlike components. Notably, the pattern features a fixed point (blue region) away from the equator of the $\mathbf{N}$-sphere. This unconventional fixed point provides the microscopic origin of deterministic perpendicular switching of $L_z$ under sufficiently strong electric fields, a mechanism absent in conventional ferromagnets~\cite{liu2012spin,miron2011perpendicular,xue2020unconventional}.

	To systematize these features, we project the first-principles torque onto an orthonormal basis of vector spherical harmonics (VSH)~\cite{Belashchenko2020,Xue2023}. The corresponding symmetry-allowed expansion in Eq.~\ref{eq:taux_full} is derived in Appendix~\ref{appendix:sot}, and it disentangles the allowed torque channels. The lowest-order even terms are $\mathrm{Re}\,\boldsymbol{Y}^{\rm D}_{1,0}\propto \hat{\mathbf{m}}\times(\hat{\mathbf{z}}\times \hat{\mathbf{m}})$ and $\mathrm{Im}\,\boldsymbol{Y}^{\rm D}_{1,1}\propto \hat{\mathbf{m}}\times(\hat{\mathbf{y}}\times \hat{\mathbf{m}})$. Both terms are staggered between sublattices and efficiently compete with the intrinsic damping $\alpha(H_{\rm E}+K)$~\cite{Xue2021}.
	To demonstrate the dynamical impact of the torque, we numerically simulate the coupled LLG equation (Eq.~\ref{eq:LLG}) using representative parameters. For the anisotropy and exchange fields we take $K=1.31$ T and $H_{\rm E}=2.06$ T, based on existing experiments and first-principles estimates~\cite{Otrokov2019,Li2022}, and we adopt a Gilbert damping parameter $\alpha=0.01$, typical for magnetic materials~\cite{Cheng2024}. The results, shown in Fig.~\ref{fig4: spindynamics}(b), reveal that a modest electric field of $0.32$ V/nm is sufficient to drive the mixed vector $\mathbf{N}$ from the equatorial plane to the opposite hemisphere within $0.6$ ns. This deterministic reversal of $L_z$, enabled entirely by the in-gap time-reversal even torque, establishes a proof of principle for efficient out-of-plane control of the Néel order without dissipative currents.
	
	Although  Fig.~\ref{fig4: spindynamics}(b) shows that the in-gap even torque alone can switch the order parameter, its magnitude $\approx 0.02 ea_0/\hbar$ is small compared to conventional metallic spin–orbit torques~\cite{Freimuth2014,Mahfouzi2018,Belashchenko2019,xue2020unconventional}, as also reflected in the chemical-potential dependence of the torkance [Fig.~\ref{fig1: crystal}(a)]. This motivates us to examine the angular dependence in the conducting regime with $\mu=-0.5$ eV, where both even and odd contributions are present and amplified. The resulting torkance distribution [Fig.~\ref{fig4: spindynamics}(c)] includes both components (individual results are provided in the Appendix~\ref{Appendix:coeffs}) and reaches a peak magnitude $\approx 1.2 ea_0/\hbar$, nearly fifty times larger than the in-gap value and comparable to metallic systems~\cite{Xue2021}. The angular pattern remains complex, with dominant dampinglike terms and a fixed point away from the equator. Corresponding spin dynamics [Fig.~\ref{fig4: spindynamics}(d)] reveal deterministic switching with a two-orders-of-magnitude reduction in critical field, $2.32$ V/$\mu$m (corresponding to $\sim 10^6$ A/cm$^2$ using reported $\sigma=300$~S/cm~\cite{Otrokov2019}), and faster response. 
	
	This enhanced performance arises from the larger torque amplitude and its predominantly dampinglike character. In the insulating regime, the in-gap torkance is dominated by time-reversal-even fieldlike components, with comparatively smaller time-reversal even dampinglike terms. Upon doping, sizable dampinglike contributions (both time-reversal even and odd) appear and dominate the switching dynamics, correlating with the much reduced threshold field in Fig.~\ref{fig4: spindynamics}(d) (see Appendix~\ref{Appendix:coeffs} for the VSH decomposition and fitted coefficients). Finally, varying the LLG parameters ($K$, $H_{\rm E}$, and $\alpha$) primarily shifts the quantitative threshold field while leaving the qualitative switching mechanism intact, underscoring its robustness. Together, these results demonstrate two regimes for N\'eel switching: (i) dissipationless in-gap torques and (ii) metallic current-induced torques with reduced thresholds. N\'eel reversal correspondingly flips the layer-resolved Chern marker and reconfigures the spin texture of the gapped edge spectrum.
	
	\section{Conclusion} \label{sec:conclusion}
	We have shown that spin–orbit torque provides an efficient mechanism to electrically control the Néel order in bilayer \ch{MnBi2Te4}. Our first-principles calculations identify a sizable dissipationless interband torque that remains finite inside the bulk gap and can deterministically switch the Néel order and the associated layer-resolved Chern marker. At finite doping, both interband and intraband contributions are amplified, reducing the critical electric field by up to two orders of magnitude. Together, these results reveal two complementary regimes for electric control: dissipationless in-gap switching without free carriers, and metallic current-driven switching with greatly reduced electric field thresholds.  
	
	Néel reversal in turn reconfigures the Berry curvature distribution between layers and reverses the spin texture of the gapped helical-like edge states. These results establish a concrete route to electrically manipulate antiferromagnetic order and its associated boundary and geometric electronic responses in a band-inverted, symmetry-indicated higher-order magnetic topological insulator.
	Experimentally, such effects could be probed via layer-resolved Hall measurements~\cite{Gao2021}, spin-torque ferromagnetic resonance~\cite{Mellnik2014}, edge transport spectroscopy~\cite{Lin2022}, and intrinsic nonlinear Hall responses~\cite{Liu2024}, which provide a direct readout of $L_z$. 
	Our results also motivate future exploration of odd-layer \ch{MnBi2Te4}, where SOT-driven switching of the local magnetization could directly toggle the quantum anomalous Hall effect, offering a pathway to reconfigurable quantized transport.

	\section{Acknowledgments} 
	The work is supported by the National Science Foundation under Grant No. OIA-2229498 and UAB startup fund. We gratefully acknowledge the resources provided by the University of Alabama at Birmingham IT-Research Computing group for high performance computing (HPC) support and CPU time on the Cheaha compute cluster. We thank Ran Cheng, Paul Haney, and Mark Stiles for fruitful discussions.
	
	\appendix
	
	\section{First-principles Methods and Topological Diagnostics}
	\label{Appendix:methods}
	\subsection{First-principles details}
	Electronic structure calculations were performed within density functional theory (DFT) using the projector-augmented-wave (PAW) method as implemented in the VASP package~\cite{VASP}. Spin–orbit coupling was included self-consistently. A plane-wave cutoff of 500 eV was used, and the optPBE-vdW functional was employed for exchange–correlation effects~\cite{perdew1996generalized}, supplemented by an on-site Hubbard correction on Mn $d$ orbitals. We used a Hubbard $U=3.9$ eV and Hund’s coupling $J_{\rm H}=0.15U$ on Mn d-orbitals~\cite{U_Mn}, which yields a local magnetic moment of $4.82~\mu_B$ per Mn. Both atomic positions and lattice parameters were fully optimized until the residual forces were below $10^{-3}$ eV/\AA{} and the total energy converged to $10^{-8}$ eV. The optimized in-plane lattice parameter is $a=b=4.361$ \AA{} and a vacuum thickness of $\sim 40$~\AA{} was used. Brillouin-zone integrations were performed on a $12\times12\times1$ $\Gamma$-centered $k$-mesh.

\begin{figure}[htbp]
		\includegraphics[width=0.9\columnwidth]{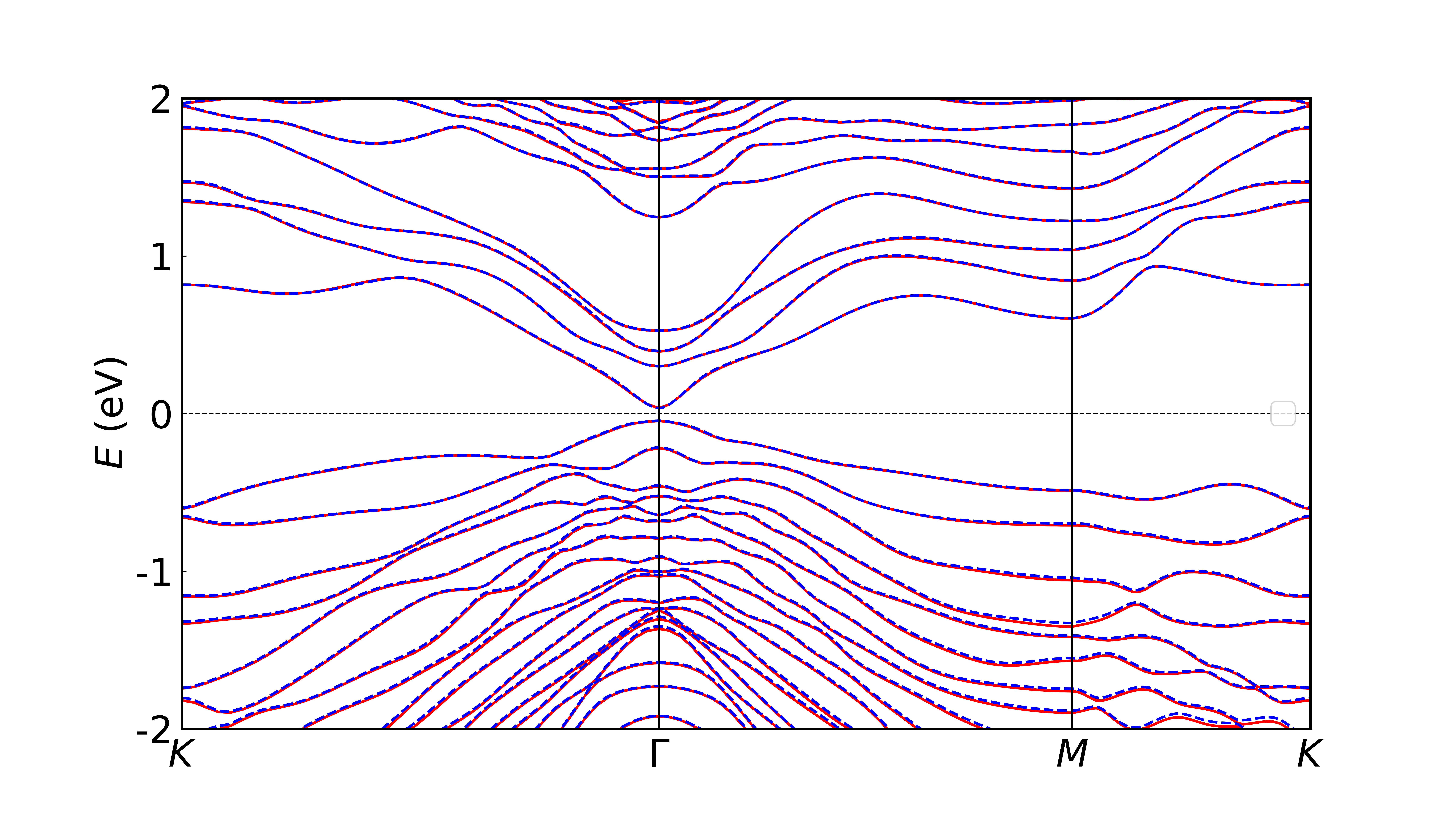}
		\caption{Band structure of bilayer \ch{MnBi2Te4} along the high-symmetry path 
			$K(\tfrac{1}{3},\tfrac{1}{3},0) \rightarrow \Gamma(0,0,0) \rightarrow M(\tfrac{1}{2},0,0) \rightarrow K(\tfrac{1}{3},\tfrac{1}{3},0)$.
			Red solid lines: VASP; blue dashed lines: final tight-binding Hamiltonian. 
			Spin degeneracy is preserved due to $\mathcal{PT}$ symmetry.}
		\label{Fig:bands}
	\end{figure}

	To analyze topological properties and electric‐field-induced linear responses, we constructed a real‐space tight‐binding Hamiltonian using Mn $d$, Bi $p$, and Te $p$ orbitals with VASP2WANNIER~\cite{mostofi2008wannier90} and Wannier90~\cite{Wannier90}.
	The Hamiltonian was further symmetrized using WannSymm~\cite{zhi2022wannsymm} to enforce the $\bar{3}m’$ magnetic point group symmetry. The resulting symmetrized tight‐binding band structure reproduces the original VASP results, as shown in Fig.~\ref{Fig:bands}.
	
	Surface spectra were obtained using the iterative Green’s function method~\cite{Greenwanniertools} implemented in WannierTools~\cite{wu2018wanniertools}. 
	\begin{figure}[htbp]
		\includegraphics[width=0.9\columnwidth]{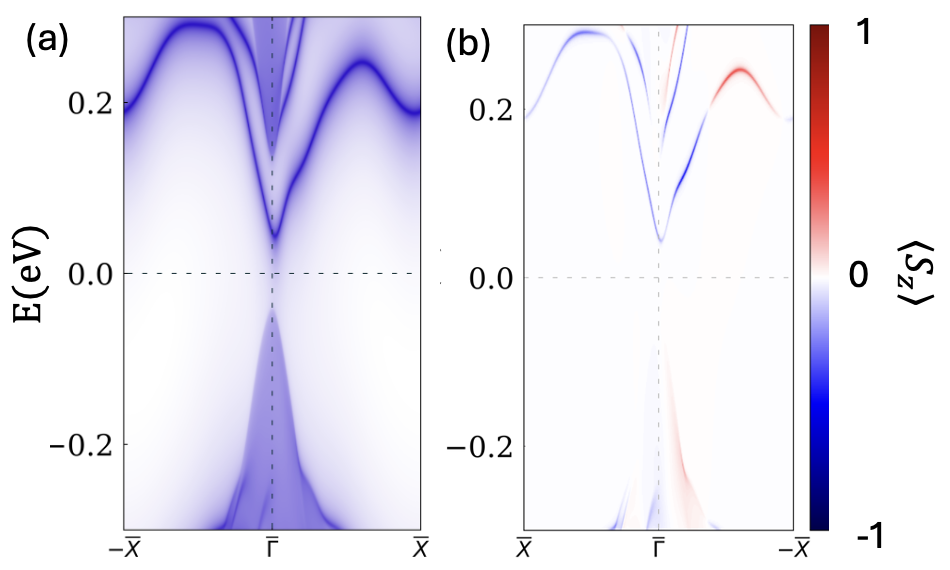}
		\caption{(a) Edge spectrum of bilayer MnBi$_2$Te$_4$ in the \emph{non-inverted} (reduced-SOC) regime, showing weak in-gap edge features.
			(b) Zoomed-in view colored by $\langle S_z\rangle$, showing no comparable spin-textured structure to the inverted case.}
		\label{Fig:edge}
	\end{figure}
	
	For completeness, we also computed the edge spectrum of bilayer MnBi$_2$Te$_4$ in the non-inverted regime (reduced SOC), as shown in Fig.~\ref{Fig:edge}. In this case, the in-gap features are weak and do not exhibit a pronounced spin-textured structure, consistent with trivial edge resonances rather than protected boundary modes. This control calculation highlights the contrast with the inverted regime discussed in the main text, where band inversion correlates with a sizable layer-resolved Chern marker and stronger spin-textured in-gap edge features.

	\subsection{Topological Characterization Methods}
	To evaluate the local Chern marker in our tight-binding model, we follow the recipe of Varnava and Vanderbilt~\cite{Vanderbilt2018},
	\begin{equation}
		C^{\rm A,B}=\frac{-4\pi }{A}\,{\rm Im}\sum_{\mathbf{k},v,v',c} 
		X_{vc\mathbf{k}}\,Y^{\dagger}_{v'c\mathbf{k}}\,\rho^{\rm A,B}_{v'v\mathbf{k}},
	\end{equation}
	where 
	\begin{equation}
		X_{vc\mathbf{k}}=\frac{\bra{\psi_{v\mathbf{k}}}i\hbar v_x\ket{\psi_{c\mathbf{k}}}}{E_{c\mathbf{k}}-E_{v\mathbf{k}}},
	\end{equation}
	the velocity matrix elements in our diagonal tight-binding approximation are
	$i\hbar \bra{j}v_x\ket{j'}=(\bar{x}_j-\bar{x}_{j'})H_{jj'}$, and $A$ is the in-plane unit cell area. 
	The sublattice projection operator is
	\begin{equation}
		\rho_{vv'\mathbf{k}}^{\rm A,B}=\sum_{z\in{\rm A,B}} 
		\psi_{v\mathbf{k}}^{*}(z)\psi_{v'\mathbf{k}}(z),
	\end{equation}
	where A and B denote orbitals localized above and below $z=0.5c$.
	This evaluation requires the full Bloch functions $\psi_{n\mathbf{k}}$ rather than only the cell-periodic parts $u_{n\mathbf{k}}$~\cite{Vanderbilt2018} because the local Chern marker depends explicitly on the real-space distribution of the wavefunctions.
	
	We evaluate the rotation topological invariants (RTIs)~\cite{Hughes2019,Murakami2021,Wang2024} from the
	$C_{3z}$ eigenvalues of the occupied bands at the rotation-invariant momenta $\Gamma$, $K=(\frac{4\pi}{3a},0)$, and $K'=-K$ using the symmetrized Wannier Hamiltonian. 
	With SOC, we use the spinful (double-group) representation of $C_{3z}$ for which
	$(C_{3z})^3=-1$, and the allowed eigenvalues are
	$\lambda_p=\exp[i\frac{\pi}{3}(2p-1)]$ with $p=1,2,3$.
	Let $\Gamma_p$ and $K_p$ ($K'_p$) denote the number of occupied bands at $\Gamma$ and $K$ ($K'$)
	with $C_{3z}$ eigenvalue $\lambda_p$. The rotation symmetry indicators are then defined as
	$[K_p^{(3)}] \equiv K_p-\Gamma_p$
	(with the redundant components removed to form a linearly independent set~\cite{Hughes2019,Murakami2021,Wang2024}).
	For the occupied manifold of the bilayer we obtain
	\begin{equation}
		\chi^{(3)}=\{[K_1^{(3)}],[K_1^{\prime(3)}],[K_2^{(3)}],[K_2^{\prime(3)}]\}
		=\{1,1,0,-1\},
	\end{equation}
	diagnosing a nontrivial $C_{3z}$-protected higher-order topological phase despite total Chern number $C=0$.

	\section{Spin-orbit torkance and vector spherical harmonics decomposition}\label{appendix:sot}
	\subsection{Linear response spin-orbit torkance}
	The spin–orbit torkance was calculated within linear-response theory. The $i^{\mathrm{th}}$ component of the torkance on sublattice A/B in response to an electric field along the $j^{\mathrm{th}}$ direction is denoted $\tau_{ij}^{\rm A,B}$. Its even and odd parts are given by~\cite{Freimuth2014,xue2020unconventional,Xue2021}
	\begin{equation}
		(\tau_{ij}^{\rm A,B})^{\rm even} = 2e\, \mathrm{Im}\sum_{n,m\neq n} f_n \frac{(\partial H/\partial k_j)_{nm} (\mathcal{T}_i^{\rm A,B})_{mn}}{(E_m - E_n)^2+\eta^2},
		\label{eq1}
	\end{equation}
	\begin{equation}
		(\tau_{ij}^{\rm A,B})^{\rm odd} = -e \sum_{n} \frac{1}{2\eta}\,\frac{\partial f_n}{\partial E_n} \, (\partial H/\partial k_j)_{nn} (\mathcal{T}_i^{\rm A,B})_{nn},
		\label{eq2}
	\end{equation}
	where $\ket{u_n}$ are eigenstates of the Bloch Hamiltonian $H_{\bf k}$ with $n=(\mathbf{k},\text{band})$. Matrix elements are $(O)_{nm}=\bra{u_n}O\ket{u_m}$. $f_n=[e^{(E_n-\mu)/k_B T}+1]^{-1}$ is the Fermi–Dirac function, $\mu$ is the chemical potential, $\eta$ is the phenomenological lifetime broadening, and $e$ is the electron charge.

	The sublattice-resolved torque operator is 
	\begin{equation}
		\mathcal{T}^{\rm A,B}=\frac{i}{2\hbar}\{[S,\Delta], P^{\rm A,B}\},
	\end{equation}
	where $S$ is the spin operator, $\Delta$ is the time-reversal odd spin-dependent exchange–correlation potential extracted from the self-consistent DFT Hamiltonian, and $P^{\rm A,B}$ projects onto orbitals centered on sites A or B.  
	
	To evaluate the torque as a function of the mixed vector $\hat{\mathbf{N}}$, we manually rotated the spins on sublattices A and B according to the $C_{2x}$ symmetry (see main text). Numerical evaluation employed a dense ${\bf k}$ mesh of $500\times500$ for Brillouin-zone integration, an angular sampling grid of $40\times80$ in $(\theta^{\rm A},\phi^{\rm A})$, a thermal broadening $k_BT=2.6$ meV (corresponding to the N\'eel temperature of 30 K), and a lifetime broadening $\eta=25$ meV (corresponding to a quasiparticle lifetime of $\sim$13 fs).
	
	\subsection{Symmetry-allowed spin-orbit torkance form}
	Following Ref.~\cite{Xue2023}, we expand the spin–orbit torque in an orthonormal basis of vector spherical harmonics (VSH). 
	For a magnetization direction $\hat{\mathbf{m}}=(\sin\theta \cos\phi,\sin\theta \sin\phi,\cos\theta)$, the VSH are defined from scalar harmonics $Y_{lm}(\hat{\mathbf{m}})$ as
	\begin{align}
		\boldsymbol{Y}^{\rm D}_{lm}(\hat{\mathbf{m}})&=\frac{\nabla_{\hat{\mathbf{m}}} Y_{lm}(\hat{\mathbf{m}})}{\sqrt{l(l+1)}},\label{eq:DampingVSH}\\
		\boldsymbol{Y}^{\rm F}_{lm}(\hat{\mathbf{m}})&=\frac{\hat{\mathbf{m}}\times\nabla_{\hat{\mathbf{m}}} Y_{lm}(\hat{\mathbf{m}})}{\sqrt{l(l+1)}}.\label{eq:FieldVSH}
	\end{align}
	Because the torque is always orthogonal to $\hat{\mathbf{m}}$, it naturally decomposes into
	dampinglike ($\boldsymbol{Y}^{\rm D}$) and fieldlike ($\boldsymbol{Y}^{\rm F}$) components.
	We label them according to their role in Landau--Lifshitz--Gilbert dynamics:
	fieldlike terms $\boldsymbol{Y}^{\rm F}_{lm}\propto \hat{\mathbf{m}}\times\nabla_{\hat{\mathbf{m}}}Y_{lm}$
	correspond to $\hat{\mathbf{m}}\times\mathbf{H}_{\rm eff}$ associated with a scalar potential
	$\mathcal{E}(\hat{\mathbf{m}})\propto Y_{lm}$ with $\mathbf{H}_{\rm eff}\propto-\nabla_{\hat{\mathbf{m}}}\mathcal{E}$,
	while dampinglike terms satisfy
	$\boldsymbol{Y}^{\rm D}_{lm}\propto \nabla_{\hat{\mathbf{m}}}Y_{lm}
	\propto \hat{\mathbf{m}}\times(\mathbf{H}_{\rm eff}\times\hat{\mathbf{m}})$.
	
	With an applied electric field $\hat{\mathbf{E}}$, the torkance takes the form
	\begin{equation}
		\boldsymbol{\tau}_{\hat{\mathbf{E}}}(\hat{\mathbf{m}})=\sum_{lm}\left[C^{\rm D}_{lm}(\hat{\mathbf{E}})\,\boldsymbol{Y}^{\rm D}_{lm}+
		C^{\rm F}_{lm}(\hat{\mathbf{E}})\,\boldsymbol{Y}^{\rm F}_{lm}\right].
	\end{equation}
	Here $C^{\rm D}_{lm}$ and $C^{\rm F}_{lm}$ quantify the contribution of each dampinglike and fieldlike channel. 
	
	Symmetry further restricts the allowed coefficients. For $\hat{\mathbf{E}}\parallel\hat{x}$, the two sublattices are related by $C_{2x}$, so it suffices to impose constraints on one sublattice. The only remaining local symmetry is the mirror $\mathcal{M}_{yz}$. From the character table in Ref.~\cite{Xue2023}, this requires $l+m$ odd (even) for real (imaginary) components of $\boldsymbol{Y}^{\rm D,F}_{lm}$.
	Applying these rules, the even and odd torkance on sublattice A reduce to:
	\begin{equation}
		\begin{split}
			&\boldsymbol{\tau}^{\text{A,even}}_{\hat{x}}=\sum_{lm}C^{{\rm F,even}}_{2l,2m+1}\,\mathrm{Re}\,\boldsymbol{Y}^{\rm F}_{2l,2m+1}
			+C^{{\rm F,even}}_{2l,2m}\,\mathrm{Im}\,\boldsymbol{Y}^{\rm F}_{2l,2m}\\
			&+C^{{\rm D,even}}_{2l+1,2m}\,\mathrm{Re}\,\boldsymbol{Y}^{\rm D}_{2l+1,2m}+C^{{\rm D,even}}_{2l+1,2m+1}\,\mathrm{Im}\,\boldsymbol{Y}^{\rm D}_{2l+1,2m+1},\\
			&\boldsymbol{\tau}^{\text{A,odd}}_{\hat{x}}=\sum_{lm}
			C^{{\rm D,odd}}_{2l,2m+1}\,\mathrm{Re}\,\boldsymbol{Y}^{\rm D}_{2l,2m+1}+
			C^{{\rm D,odd}}_{2l,2m}\,\mathrm{Im}\,\boldsymbol{Y}^{\rm D}_{2l,2m}\\
			&+C^{{\rm F,odd}}_{2l+1,2m}\,\mathrm{Re}\,\boldsymbol{Y}^{\rm F}_{2l+1,2m}+C^{{\rm F,odd}}_{2l+1,2m+1}\,\mathrm{Im}\,\boldsymbol{Y}^{\rm F}_{2l+1,2m+1},
		\end{split}\label{eq:taux_full}
	\end{equation}
	with $l\geq0$ and $0\leq m\leq l$. In practice, we expand the first-principles torque up to $l=15$. Since the VSH form a complete orthonormal basis, the expansion converges systematically, and the fitted results are insensitive to the cutoff. We have verified that the key features discussed in the main text, such as the staggered components and the existence of a fixed point away from the equator, are already captured at low orders, and remain unchanged upon extending the expansion to $l=15$. 
	
	A notable outcome is the appearance of the lowest-order dampinglike term $\mathrm{Re}\,\boldsymbol{Y}^{\rm D}_{1,0}\propto \hat{\mathbf{m}}\times(\hat{z}\times \hat{\mathbf{m}})$ in both \ch{CrI3}~\cite{Xue2021} and \ch{MnBi2Te4}. In bilayer \ch{CrI3}, this is expected since no threefold rotation axis exists. By contrast, in \ch{MnBi2Te4} one might expect this term to be forbidden by the $C_{3z}$ rotation of each layer, as in monolayer \ch{Fe3GeTe2}~\cite{Xue2023}. In the reduced mixed vector space $\mathbf{N}$, however, the $C_{3z}$ constraint is effectively lifted, rendering the term allowed. This underscores that the accessible torque channels depend on whether symmetry is enforced in Néel space or in the mixed vector representation.

	\subsection{Symmetry distinction between mixed-vector space and N\'eel space}
	\begin{figure}[htbp]
		\includegraphics[width=1\linewidth]{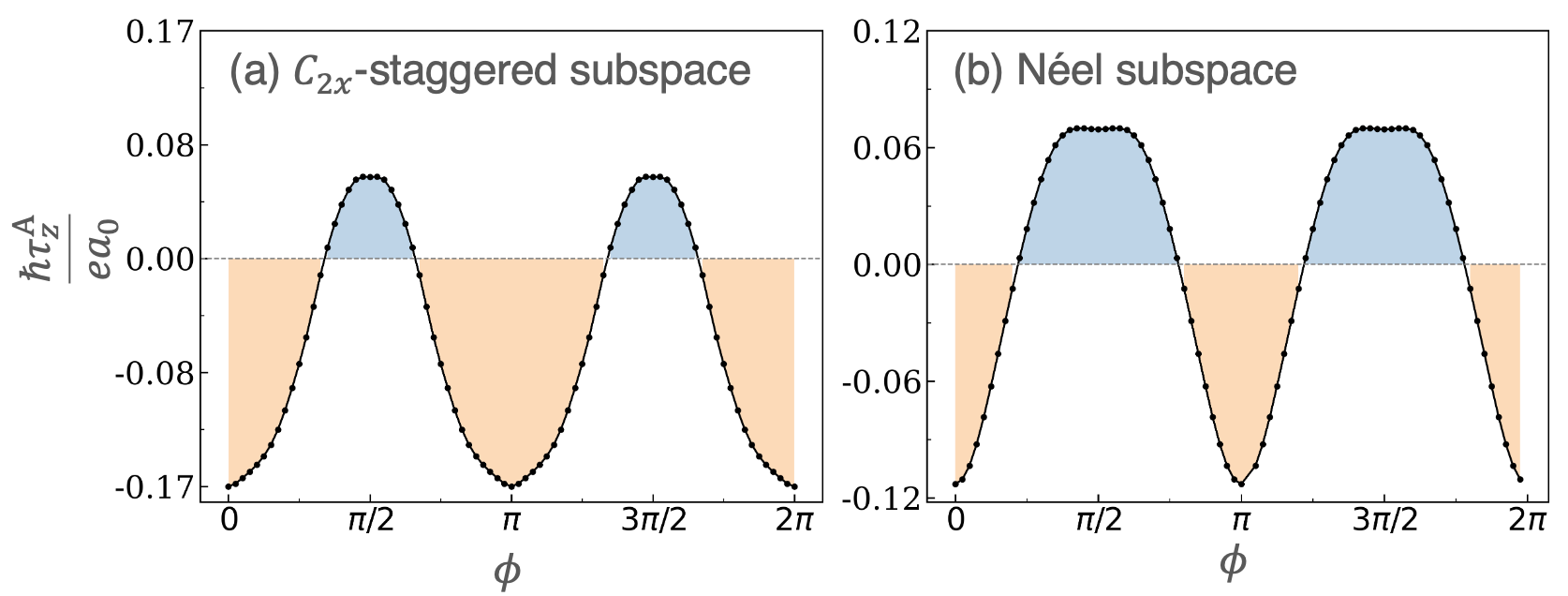}
		\caption{Out-of-plane torkance $\tau_z$ as a function of the azimuthal angle $\phi$ at the equator ($\theta=\pi/2$).  
			(a) In the $C_{2x}$-staggered subspace the torkance has a nonzero mean value.  
			(b) In the Néel subspace the mean torkance vanishes.  
			Shaded regions denote positive (blue) and negative (orange) $\tau_z$.  
			Both curves are plotted for a chemical potential $\mu = -0.3$ eV below the valence band maximum.}
		\label{Fig:equator}
	\end{figure}
	
	The bilayer \ch{MnBi2Te4} crystal possesses a threefold rotation symmetry about the $z$ axis ($C_{3z}$). However, if we restrict to the twofold-staggered subspace (the “$C_{2x}$-staggered” subspace),
	\begin{equation}
		\theta^B = \pi + \theta^A,\qquad
		\phi^B   = \pi - \phi^A,
	\end{equation}
	then $C_{3z}$ is no longer preserved. Under a $C_{3z}$ rotation,
	\begin{align}
		(\theta^A,\phi^A)&\xrightarrow{C_{3z}}(\theta^A,\;\phi^A+2\pi/3),\\
		(\theta^B,\phi^B)&\xrightarrow{C_{3z}}(\pi+\theta^A,\;\pi-\phi^A+2\pi/3),
	\end{align}
	which violates the defining $C_{2x}$ constraint since
	\[
	\pi-\phi^A+2\pi/3 \;\neq\; \pi-(\phi^A+2\pi/3).
	\]
	By contrast, in the collinear Néel subspace (where only $\theta^B=\pi+\theta^A$), $C_{3z}$ rotations map allowed states onto themselves.  
	
	This distinction has a direct implication for the vector spherical harmonics expansion. In the Néel subspace, $C_{3z}$ symmetry enforces that only coefficients with $m \pmod{3} \neq 0$ survive. In the staggered subspace, the effective breaking of $C_{3z}$ lifts the rotational constraint, permitting finite $m=0$ terms at all chemical potentials. 
	As shown in Fig.~\ref{Fig:mu dependence}, these coefficients remain nonzero across the entire range. 
	This corresponds to a constant out-of-plane component of the torkance at $\theta=\pi/2$, visible in Fig.~\ref{Fig:equator}. 
	By contrast, in the Néel subspace $m=0$ terms are symmetry-forbidden, and the mean $\tau_z$ at the equator must vanish. 
	This distinction provides an experimental signature: the presence of a finite equatorial $\tau_z$ signals staggered-subspace dynamics, while its absence indicates confinement to the Néel subspace.

	\section{Chemical-potential dependence of Spin-orbit torkance} \label{Appendix:coeffs}
	
	Following the recipe described in the main text, we report here the fitted vector spherical harmonic (VSH) coefficients of the spin–orbit torkance and illustrate their angular and chemical–potential dependence.
	
	\begin{table}[htbp]
		\centering
		\begin{tabular}{|c|c|c|c|c|}
			\hline
			$C^{\rm D,even}_{1,1}$ & $C^{\rm F,even}_{2,1}$ & $C^{\rm F,even}_{4,1}$ & $C^{\rm F,even}_{2,2}$ & $C^{\rm F,even}_{4,2}$ \\
			\hline
			$-0.0094$ & $-0.0668$ & $0.0071$ & $-0.0206$ & $-0.0051$ \\
			\hline
		\end{tabular}
		\caption{Leading VSH coefficients for the time-reversal-even (interband) torkance on sublattice A in the gap ($\mathbf{E}\parallel\hat{x}$). Coefficients are in units of $ea_0/\hbar$; terms with $|C|<5\times10^{-3}$ are omitted.}
		\label{tab:vsh_fit_updated}
	\end{table}
	
	Table~\ref{tab:vsh_fit_updated} lists the leading expansion coefficients of the in-gap even torkance. In the in-gap regime, higher-order torque components such as the large fieldlike $C^{\rm F,even}_{2,1}$ term exceed the lowest-order dampinglike $C^{\rm D,even}_{1,1}$ contribution.

	\begin{table}[htbp]
		\centering
		\begin{tabular}{|c|c|c|c|c|c|c|}
			\hline
			$C^{\rm D,even}_{3,2}$ & $C^{\rm D,even}_{1,1}$ & $C^{\rm D,even}_{3,3}$ & 
			$C^{\rm F,even}_{2,1}$ & $C^{\rm F,even}_{4,3}$ & 
			$C^{\rm F,even}_{2,2}$ & $C^{\rm F,even}_{4,2}$ \\
			\hline
			$0.10$ & $-2.29$ & $-0.36$ & $0.46$ & $0.16$ & $0.21$ & $-0.20$ \\
			\hline
		\end{tabular}
		\vspace{0.2cm}
		
		\begin{tabular}{|c|c|c|c|c|c|}
			\hline
			$C^{\rm D,odd}_{2,1}$ & $C^{\rm F,odd}_{1,0}$ & $C^{\rm F,odd}_{1,1}$ & 
			$C^{\rm F,odd}_{3,1}$ & $C^{\rm F,odd}_{3,3}$ & $C^{\rm F,odd}_{5,5}$ \\
			\hline
			$-0.90$ & $0.15$ & $1.50$ & $0.16$ & $-0.37$ & $-0.14$ \\
			\hline
		\end{tabular}
		
		\caption{Leading VSH coefficients for the fitted torkance on sublattice A at $\mu=-0.5$~eV ($\mathbf{E}\parallel\hat{x}$). Top: time-reversal-even (interband); bottom: time-reversal-odd (intraband). Coefficients are in units of $ea_0/\hbar$; terms with $|C|<0.01$ are omitted.}
		\label{tab:vsh_fit_even_odd_mu-0.5}
	\end{table}

	Tables~\ref{tab:vsh_fit_even_odd_mu-0.5} list the leading expansion coefficients of the spin–orbit torkance at $\mu=-0.5$ eV. While the lowest-order terms such as $\mathrm{Re}\boldsymbol{Y}^{\rm D,F}_{1,0}$ and $\mathrm{Im}\boldsymbol{Y}^{\rm D,F}_{1,1}$ appear, additional higher-order components are also present. In particular, the sizable odd dampinglike $C^{\rm D,odd}_{2,1}\mathrm{Re}\boldsymbol{Y}^{\rm D}_{2,1}$ term is important for reducing the critical field for Néel-order switching. These fitted coefficients serve as inputs for the LLG simulations presented in the main text.

	\begin{figure}[htbp]
		\includegraphics[width=0.95\linewidth]{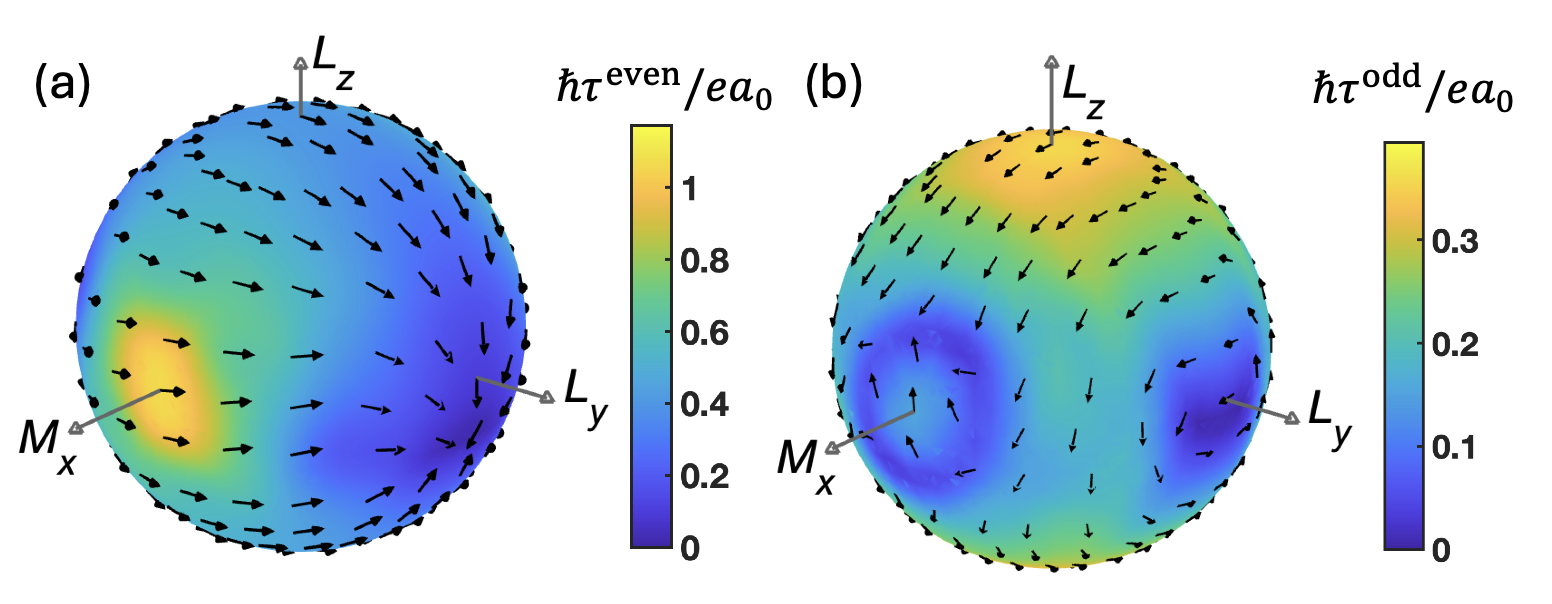}
		\caption{Angular dependence of the spin–orbit torkance at $\mu=-0.5$ eV. 
			(a) Time-reversal–even (interband) component $\hbar\tau^{\rm even}/ea_0$. 
			(b) Time-reversal–odd (intraband) component $\hbar\tau^{\rm odd}/ea_0$. 
			Arrows denote the torque direction on the mixed-vector sphere; color indicates the magnitude.}
		\label{fig: torkance}
	\end{figure}
	
	Figure~\ref{fig: torkance} visualizes the angular dependence reconstructed from these coefficients. The even (interband) contribution is generally larger in magnitude, whereas the odd (intraband) part is more pronounced near the easy axis.

	\subsection{Chemical-potential dependence and experimental relevance}
	\begin{figure}[htbp]
		\includegraphics[width=1\columnwidth]{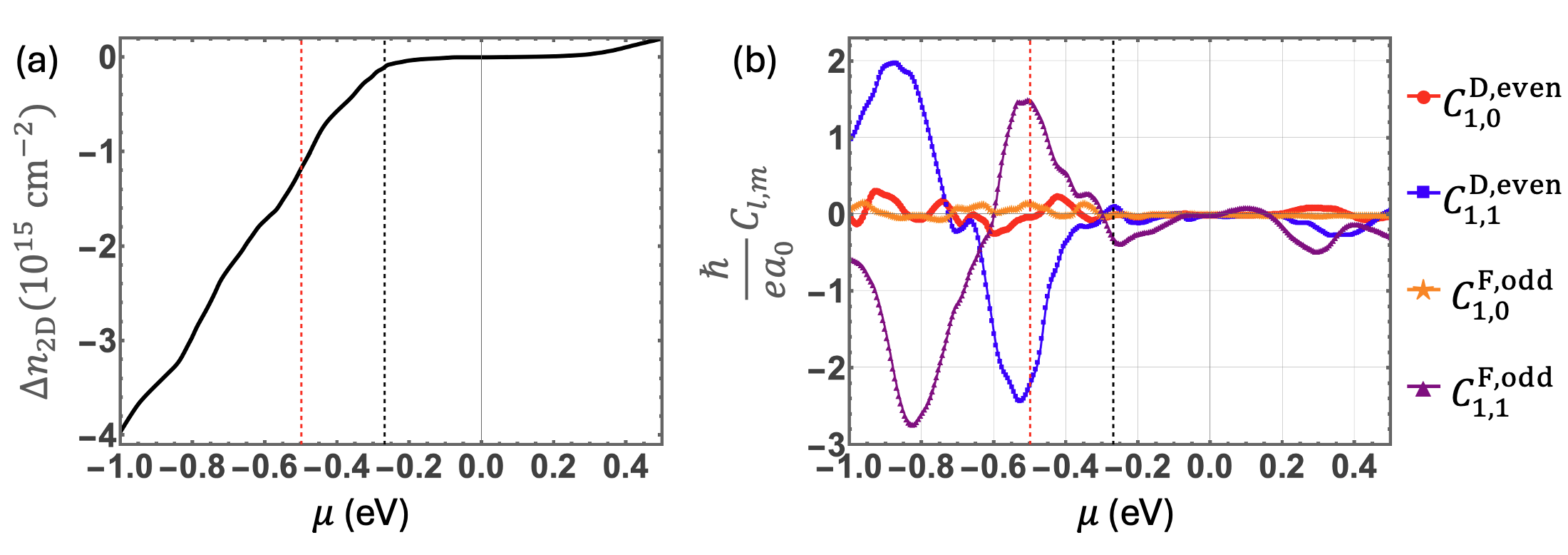}
		\caption{
			(a) Carrier sheet density change $\Delta n_{2\rm D}$ as a function of chemical potential $\mu$, obtained from the DFT density of states using the in-plane unit-cell area ($a=4.361$~\AA). 
			(b) Lowest-order even components ($\mathrm{Re}\,\boldsymbol{Y}^{\rm D}_{1,0}$ and $\mathrm{Im}\,\boldsymbol{Y}^{\rm D}_{1,1}$) and odd components ($\mathrm{Re}\,\boldsymbol{Y}^{\rm F}_{1,0}$ and $\mathrm{Im}\,\boldsymbol{Y}^{\rm F}_{1,1}$) of the torkance expansion for $\mathbf{E}\parallel\hat{x}$. 
			The red and black vertical dashed lines mark $\mu=-0.5$ eV and $\mu=-0.27$ eV, respectively, corresponding to the representative switching regimes discussed in the text.  Calculations use a lifetime broadening $\eta=25$ meV.}
		\label{Fig:mu dependence}
	\end{figure}
	
	In the insulating (in-gap) regime, the predicted switching threshold 
	$E_c \approx 0.32$ V/nm represents an upper-bound case. 
	At such large fields, intrinsic interband breakdown processes 
	(e.g., Zener-type tunneling or contact-assisted breakdown) 
	may become relevant. We therefore regard the insulating limit primarily 
	as establishing the symmetry-enforced existence and functional form 
	of the interband (time-reversal–even) torque. 
	In experimental implementations, this regime would most naturally 
	be accessed using short electrical pulses to mitigate heating 
	and extrinsic breakdown effects.
	
	Bilayer \ch{MnBi2Te4} is typically $n$-doped due to native defects and is readily
	tunable by electrostatic gating~\cite{Zeugner2019,Otrokov2019,Deng2020}. To connect
	the chemical potential $\mu$ to an experimentally measurable carrier density, we
	compute the change in sheet density relative to the charge-neutral point,
	\begin{equation}
		\Delta n_{\rm 2D}(\mu)=\frac{1}{A_{\rm uc}}
		\sum_{n,\mathbf{k}} \left[ f_{n\mathbf{k}}(\mu)-f_{n\mathbf{k}}(\mu_0) \right],
	\end{equation}
	where $A_{\rm uc}=a^2\sqrt{3}/2$ is the in-plane unit-cell area ($a=4.361$~\AA)
	and $\mu_0$ denotes the Fermi level at charge neutrality; throughout we measure
	$\mu$ relative to $\mu_0$. Figure~\ref{Fig:mu dependence}(a) shows that
	$\Delta n_{2\rm D}$ varies nonlinearly with $\mu$ due to the energy-dependent
	density of states. The representative value $\mu=-0.5$~eV corresponds to
	$\Delta n_{2\rm D}\sim 10^{15}$~cm$^{-2}$, whereas a more moderate shift
	$\mu\approx -0.27$~eV corresponds to $\Delta n_{2\rm D}\sim 10^{14}$~cm$^{-2}$,
	achievable under strong electrostatic or ionic gating.
	
	Importantly, the switching behavior is not monotonic in $\mu$. Changing $\mu$
	modifies not only the overall torque magnitude but also its angular composition
	(i.e., the relative weight of symmetry-allowed VSH channels), which shifts the
	mixed-vector fixed points and therefore the qualitative dynamics. This is
	illustrated in Fig.~\ref{Fig:mu dependence}(b): the leading even/odd
	coefficients evolve strongly and non-monotonically once $\mu$ enters the
	conducting regime, providing the microscopic basis for distinct dynamical outcomes at different dopings.
	
\begin{figure}[htbp]
		\includegraphics[width=1\columnwidth]{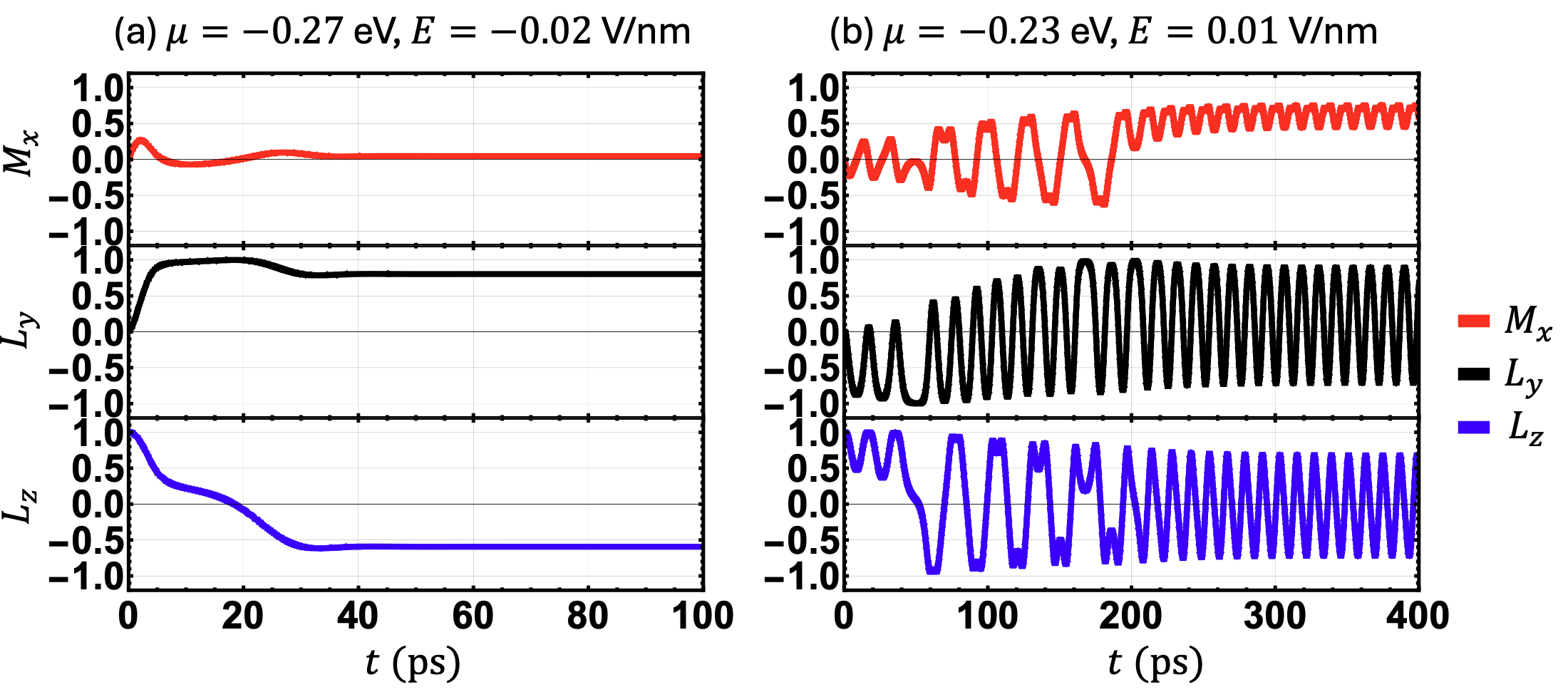}
		\caption{Representative Landau--Lifshitz--Gilbert dynamics of the mixed-vector
			components $\mathbf{N}=(M_x,L_y,L_z)$ at two chemical potentials.
			(a) At $\mu=-0.27$~eV, an in-plane field $E=-0.02$~V/nm drives deterministic
			reversal of the N\'eel component $L_z$ on a sub-100~ps timescale.
			(b) At $\mu=-0.23$~eV and $E=0.01$~V/nm, the dynamics enter a sustained
			oscillatory (precessional) regime with no steady N\'eel reversal, illustrating
			the non-monotonic dependence of the switching behavior on $\mu$ through changes
			in the torque amplitude and angular composition.}
		\label{Fig:LLG}
	\end{figure}

	As a concrete moderate-doping example, at $\mu\approx -0.27$~eV we obtain deterministic N\'eel reversal with a reduced critical field
	$|E_c|\approx 0.02$~V/nm (Fig.~\ref{Fig:LLG} (a)), more than an order of magnitude smaller than the insulating limit. Using experimentally reported conductivities $\sigma \sim 300$--$1000$~S/cm for \ch{MnBi2Te4}~\cite{Zeugner2019,Otrokov2019},
	this corresponds to current densities $J=\sigma E\sim 10^{6}$--$10^{7}$~A/cm$^2$, which fall within the range commonly used for deterministic SOT switching in thin-film devices~\cite{Chang2024,Fan2014,Han2017}. 
	For intermediate $\mu$ (e.g., $\mu\approx -0.23$~eV), the altered balance between torque channels can instead produce sustained oscillatory (precessional) trajectories without a steady N\'eel reversal, as shown in Fig.~\ref{Fig:LLG} (b).  Similar SOT-driven oscillatory dynamics were reported in bilayer \ch{CrI3}~\cite{Xue2021}.
	
	We emphasize that (i) $\mu=-0.5$~eV serves as an optimized example highlighting the enhancement mechanism; (ii) moderate carrier densities already yield a substantial reduction of the switching threshold; and (iii) the critical field further scales with the anisotropy, exchange, and damping parameters entering the LLG dynamics, providing additional pathways for device-level optimization.
	Overall, while the insulating limit establishes the symmetry-driven interband mechanism, the experimentally relevant doped regime yields thresholds compatible with established transport conditions.
	
	\subsection{Robustness to weak mirror-symmetry breaking}
	The main text analysis assumes the ideal bilayer symmetry in which the in-plane mirror $\mathcal{M}_{yz}$ imposes selection rules on the VSH decomposition, forbidding specific harmonics in the torque expansion. In realistic devices, substrate, gating, strain, or defects can weakly break $\mathcal{M}_{yz}$, thereby allowing additional VSH components that are symmetry-forbidden in the ideal limit. To assess robustness, we explicitly include the leading 
	$\mathcal{M}_{yz}$-forbidden harmonic 
	$\epsilon\,|C^{\rm F,even}_{2,1}|\,\mathrm{Re}\,\boldsymbol{Y}^{\rm D}_{1,1}$ 
	in the even (interband) torque on each sublattice, where $\epsilon$ 
	controls the strength relative to the dominant fitted in-gap coefficient 
	$C^{\rm F,even}_{2,1}$ (Table~\ref{tab:vsh_fit_updated}). 
	We then repeat the sublattice-resolved LLG simulations.
	We find that deterministic N\'eel reversal persists for all tested $|\epsilon|\le 0.05$; mirror breaking produces only quantitative shifts of fixed points and modest changes of the threshold field, which can increase or decrease depending on the sign of $\epsilon$ (see Table~\ref{tab:eps_robust}).
	\begin{table}[htbp]
		\centering
		\begin{tabular}{|c|c|c|c|}
			\hline
			$\epsilon$ & $E_c$ (V/nm) & $L_z(t_{\rm end})$ & Switching \\
			\hline
			$-0.05$ & 0.24 & -0.895 & Yes \\
			$-0.02$ & 0.27 & -0.872 & Yes \\
			$0$ & 0.32 & -0.834 & Yes \\
			$+0.02$ & 0.37 & -0.378 & Yes \\
			$+0.05$ & 0.57 & -0.443 & Yes \\
			\hline 
		\end{tabular}
		\caption{Robustness of deterministic switching against weak mirror-symmetry breaking modeled by adding the leading mirror-forbidden harmonic to the even (interband) torque with amplitude $\epsilon$ (normalized to $|C^{\rm F,even}_{2,1}|$). Switching is defined by $L_z$ crossing equator during the pulse; $L_z(t_{\rm end})$ is reported at the end of the pulse before full anisotropy-driven relaxation. After the pulse, the easy-axis anisotropy relaxes trajectories in the $L_z<0$ basin to the south pole.}
		\label{tab:eps_robust}
	\end{table}

	\section{Band-inversion-enhanced in-gap torkance in a two-band ferromagnetic insulator}
	\label{Appendix:model}
\begin{figure}[htbp]
		\centering
		\includegraphics[width = 1\linewidth]{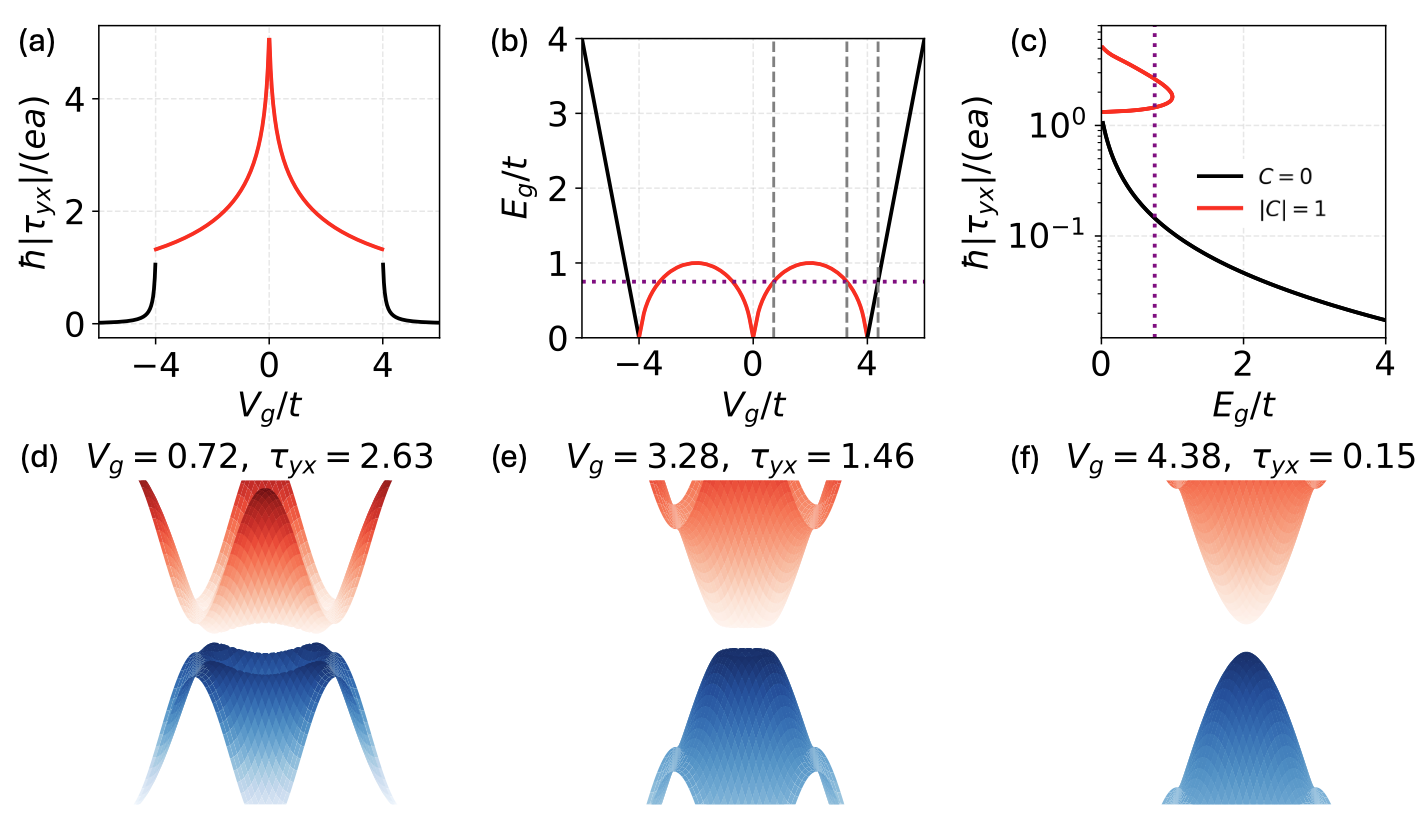}
		\caption{(a) In-gap torkance magnitude $|\tau_{yx}|$ (in units of $ea/\hbar$, with $a=1$) versus gate parameter $V_g$ (in units of hopping $t=1$). Sharp changes indicate band gap closing and transitions between distinct Chern sectors.
			(b) Band gap $E_g$ versus $V_g$. The horizontal purple dotted line marks a fixed gap size $E_g=0.75t$, realized at three distinct $V_g$ values (gray vertical dashed lines): two in the Chern insulating phase ($C=+1$) and one in the trivial phase ($C=0$).  An additional three solutions exist on the negative $V_g$ side by symmetry, yielding identical torque magnitudes; these are omitted for clarity.
			(c) Torque $\tau_{yx}$ versus band gap $E_g$. For the same $E_g$, distinct torque values appear depending on the band inversion: non-inverted/trivial ($C=0$, black) versus inverted/Chern ($|C|=1$, red).
			(d–f) Band dispersions at the three representative $V_g$ values ($0.72t,~3.28t,~4.38t$) with identical gap size $E_g=0.75t$, demonstrating the large variation in torque magnitude between phases [$\hbar|\tau_{yx}|/(ea)=2.63,~1.46,~0.15$]. Parameters: $a=1$, $t=1$, $M_z=-4$, $v_F=0.5$.}
		\label{fig:toymodel}
	\end{figure}

	We begin with a two-band lattice model on a square lattice with out-of-plane magnetization $M_z$,
	a gate-tunable mass term $V_g$, and a Rashba term generated by broken inversion symmetry along $\hat{z}$~\cite{Qi2006}.
	While this model is not intended to reproduce the full multiband electronic structure of MnBi$_2$Te$_4$,
	it captures the same essential ingredient underlying interband torques: regions in ${\bf k}$-space with
	small \emph{direct} gaps that enhance Kubo interband denominators.
	
	\begin{equation}
		\label{eq:H of FM}
		H = \Delta\sigma_z+v_F (\boldsymbol{d}\times\boldsymbol{\sigma})\cdot\hat{z},
	\end{equation}
	where $\Delta = M_z + V_g + 2t (2 - \cos{k_xa} - \cos{k_ya})$, $\boldsymbol{d} = (\sin{k_xa}, \sin{k_ya})$, and $\boldsymbol{\sigma} = (\sigma_x,\sigma_y,\sigma_z)$ are Pauli matrices. 
	This minimal model describes a ferromagnetic insulator whose band gap is controlled by the interplay of $M_z$ and $V_g$. 
	Varying $V_g$ tunes the Dirac mass and drives transitions between distinct Chern sectors:
	as shown in Fig.~\ref{fig:toymodel}(a), the system alternates between a Chern-insulating phase ($C=\pm1$)
	and a trivial insulator ($C=0$).
	
	The critical phase boundaries~\cite{Qi2006} occur at $V’_g \equiv V_g+M_z = 0,~4t,~8t$.
	Since we set $M_z=-4t$ in our calculation, these transitions appear at $V_g=-4t,0,$ and $4t$, consistent with the gap evolution shown in Fig.~\ref{fig:toymodel}(b).
	
	The torque operator is defined as
	\begin{equation}
		\boldsymbol{T} = \dot{\boldsymbol{m}} 
		= \frac{i}{\hbar}[H,\boldsymbol{m}]
		= -\frac{i}{\hbar}[M_z \sigma_z,\boldsymbol{\sigma}]
		= \frac{2M_z}{\hbar}\,(\boldsymbol{\sigma}\times \hat{z}).
	\end{equation}
	In equilibrium the spin expectation value aligns with $\hat{\boldsymbol{m}}$, 
	\begin{equation}
		\langle \psi_{v,c}|\boldsymbol{\sigma}/2|\psi_{v,c}\rangle=\mp \hat{\boldsymbol{m}}/2,
	\end{equation}
	so the net torque vanishes in equilibrium as expected.  
	
	To capture the nonequilibrium response to an applied electric field $\boldsymbol{E}=E\hat{x}$, we evaluate the interband and intraband contributions within the Kubo formalism for a generic operator $\mathcal{O}$:  
	\begin{equation}
		\label{eq:inter}
		\langle \mathcal{O}\rangle^{\rm inter}
		= \frac{2e}{\hbar}\,\mathrm{Im}\,
		\frac{\langle \psi_v|\partial H/\partial k_x|\psi_c\rangle 
			\langle \psi_c|\mathcal{O}|\psi_v\rangle}{4E_k^2},
	\end{equation}
	\begin{equation}
		\langle \mathcal{O}\rangle^{\rm intra}
		= -\frac{e}{2\eta}\,\frac{\partial f}{\partial E}\,
		\mathrm{Re}\,\langle \psi_v|\partial H/\partial k_x|\psi_v\rangle
		\langle \psi_v|\mathcal{O}|\psi_v\rangle,
	\end{equation}
	with $E_k=\sqrt{\Delta^2+v_F^2d^2}$. 
	Because the system is insulating, $\partial f/\partial E=0$ within the gap throughout the Brillouin zone, so the intraband contribution vanishes. This reflects the fact that dissipative, time-reversal–odd processes are forbidden in the insulating state.
	
	Figure~\ref{fig:toymodel}(a) plots the symmetry-allowed torque $\tau_{yx}$ versus $V_g$. It is finite in both the trivial (black) and Chern (red) phases, but is strongly enhanced in the latter. Panel (c) shows $\tau_{yx}$ versus the band gap: for a fixed gap (dotted line), the Chern phase exhibits torques nearly an order of magnitude larger than the trivial phase. The dispersions in Figs.~\ref{fig:toymodel}(d)–(f) illustrate that identical gap sizes can correspond to distinct topologies, yielding sharply different torque responses.
	Although the \emph{global} band gap $E_g$ can be identical, the inverted (Chern) regime typically contains a larger
	phase-space volume of ${\bf k}$-points with small \emph{direct} gaps (interband ``hot spots''),
	whereas in the trivial regime such near-degeneracies occur only at isolated points.
	Because the interband contribution scales as $(E_c-E_v)^{-2}$, this enlarged hot-spot phase space strongly
	amplifies the in-gap torkance in the inverted regime.

	Having illustrated the mechanism in this minimal model, we now generalize Eq.~\ref{eq:H of FM} to a four-band model describing a $\mathcal{PT}$-symmetric antiferromagnet with staggered inversion-symmetry breaking:
	\begin{equation}
		H = \Big(\Delta\sigma_z+v_F (\boldsymbol{d}\times\boldsymbol{\sigma})\cdot\hat{z}\Big)\tau_z
		+ A \tau_x ,
	\end{equation}
	where $\tau_{x,y,z}$ are Pauli matrices in the layer pseudospin space.
	In the absence of interlayer tunneling ($A=0$), the system reduces to two decoupled ferromagnetic layers (Eq.~\ref{eq:H of FM}) related by $\mathcal{PT}$ symmetry. In this limit, only time-reversal–even torques are symmetry-allowed, appearing with opposite signs on the two layers:
	\begin{equation}
		\langle T \rangle^{\rm t/b} = \pm v_F \,\boldsymbol{m}\times (\hat{y}\times \boldsymbol{m}) .
	\end{equation}
	The layer-resolved Chern markers are likewise staggered. 
	Including interlayer tunneling ($A\neq 0$) perturbatively does not
	qualitatively alter this result, leaving both the staggered torque
	and the layer-resolved Chern marker intact.

	\bibliography{reference.bib}
	
\end{document}